\documentclass[preprint2]{aastex} 

\begin{document}

\title{Absolute--Magnitude Distributions of Supernovae}  

\author{Dean Richardson\altaffilmark{1}, 
Robert L. Jenkins III\altaffilmark{2}, John Wright\altaffilmark{1}, Larry Maddox\altaffilmark{3,4}  \\ 
(drichar7@xula.edu)} 

\altaffiltext{1}{Dept. of Physics, Xavier University of Louisiana, New Orleans, LA 70125 }  
\altaffiltext{2}{Applied Physics Dept., Richard Stockton College, Galloway, NJ 08205 } 
\altaffiltext{3}{Dept. of Chemistry \& Physics, Southeastern Louisiana University, Hammond, LA 70402 }  
\altaffiltext{4}{Current address: Northrop Grumman Corporation, Oklahoma City 73135 }

\begin{abstract} 
The absolute--magnitude distributions of seven supernova types are presented. The data used  
here were primarily taken from the Asiago Supernova Catalogue, but were supplemented with additional data. 
We accounted for both foreground and host--galaxy extinction. 
A bootstrap method is used to correct the samples for Malmquist bias. 
Separately, we generate volume--limited samples, restricted to events within 100 Mpc.  
We find that the superluminous events (M$_B < -21$) make up only about 0.1\% of all supernovae in the 
bias--corrected sample. The subluminous events (M$_B > -15$) make up about 3\%.    
The normal Ia distribution was the brightest with a mean absolute blue magnitude of $-$19.25. The IIP 
distribution was the dimmest at $-$16.75.

\end{abstract} 


\section{Introduction} 
About two decades ago, a study was carried out on the absolute--magnitude distributions 
of supernovae (SNe) separated by types \citep[][hereafter MB90]{miller90}. This study was based 
on data taken from the Asiago Supernova Catalogue \citep[the ASC;][]{barbon89}, 
which at the time had 687 SNe. A decade later \citet[][hereafter R02]{richardson02} 
carried out a similar study; however, by that time the ASC had grown to 1910 SNe.  
Currently, the ASC has over 6100 SNe. A histogram of the number of supernovae versus the 
year of discovery is shown in Figure 1. In this graph we can see a drastic increase 
in the discovery rate of SNe over the last 20 years. Also shown in this figure are 
labels for important milestones in astronomy that had an effect on the discovery rate.  
The years for the studies MB90 and R02 are shown for context.  
The inset shows the same data on a log scale. Here we plot log of N+1 so that years with 
only one SN discovery can be shown. 
The large spike in the number of SNe discovered over the last decade  
is primarily due to the number of large searches that have taken place during that time.  
Many of these searches were motivated by the discovery of the accelerating universe 
\citep{riess98,perlmutter99}.  

Because of this drastic increase in the overall number of SNe, it is time 
to update the absolute--magnitude distributions for the various SN types.  
Absolute--magnitude distributions are important in determining SN rates and providing 
information useful for constraining progenitor and explosion models. 
They are also useful for the planning of future SN searches. 

The main push for the large SN searches is the discovery of type Ia SNe, which are 
used as cosmological distance indicators. As a by--product of this, more SNe of all 
types are being discovered, although follow up is not necessarily done.  
Many studies have looked at SNe~Ia in great detail and through some form of light--curve 
standardization have reduced the dispersion to about 0.12 \citep{bailey09}.   
This type of standardization is not possible with our large data set and is not carried out here. 
Most of the data used in this study are taken from the ASC. However, a number of other sources 
are used as well.  

The data used in this study, and their sources, are discussed in Section~2. The different types 
of analysis are discussed in Section~3. The results of our analysis are presented in Section~4 
along with comparisons to other studies. A summary is given in Section~5.

\section{Data} 
The primary source for our data is the ASC. This catalog contains all SNe reported with an IAU 
designation. 
In the ASC, there are more peak apparent magnitudes in the B--band than any other band.  
It is for this reason we chose to work with B magnitudes. 
In order to determine the absolute magnitudes we collected data for apparent magnitude, distance, 
foreground extinction and host--galaxy extinction. The apparent magnitudes 
were often not given in the B--band and therefore required K--corrections.
Distances were determined by several methods. 
Host--galaxy extinction estimates were made for as many SNe as possible based on published data.  

Even though most of the apparent magnitude data were taken from the ASC, there are a number of cases where 
other data were used. The bulk of the additional data were taken from \cite{li11}, \cite{kowalski08} 
and \cite{perlmutter99}. Apparent magnitudes from \cite{perlmutter99} already include K--corrections.    
Peak apparent magnitudes were also taken from \cite{stoll11}, \cite{chatzopoulos11}, 
\cite{inserra12a}, \cite{yuan10}, \cite{bassett07}, \cite{miller10}, \cite{kowal71}, 
\cite{malesani04}, \cite{drake09}, \cite{kleiser11}, \cite{cano11}, 
\cite{richmond12}, \cite{leloudas12}, \cite{clocchiatti00}, \cite{matheson00},   
\cite{pastorello10}, \cite{smith08}, \cite{rest11}, \cite{inserra12b}, \cite{pastorello02}, 
\cite{bembrick02}, \cite{ergon14}, \cite{galyam09}, and \cite{eliasrosa06}.  
For a few SNe, changes to redshift and SN type were made using the following sources: 
\cite{quimby07b}, \cite{niemela85}, \citet{benetti98,benetti99}, \cite{smith09}, 
\cite{riess99}, \cite{taddia12}, and \cite{pastorello12}.

\subsection{Apparent Magnitudes} 
Because the apparent magnitude data comes from a wide range of sources, we do not have uncertainties 
for each SN. We estimate the observational uncertainty in apparent magnitudes to be 
0.1 magnitudes. While the newer SNe 
tend to have smaller uncertainties than this, the older ones on average have larger uncertainties.   
Because apparent magnitudes were not always reported in the B--band, it was necessary to 
approximate the K--corrections for each SN. A spectrum at peak was not available for all SNe,  
therefore, we used a small set of available spectra for each SN type. We calculated a K--correction 
for each SN using each spectrum of that type. We therefore had multiple K--corrections for each SN 
and used the average for that SN. For some SN types, the number of spectra used is quite low. 
This is due to the low number of publicly available spectra at peak magnitude and with the 
necessary wavelength coverage for those SN types. There were, however, cases where certain 
spectra needed to be extended in the IR or UV. This was done with an appropriate black-body curve.    
For SNe~Ia, spectra were used from SNe 1981B \citep{branch83}, 1996X \citep{salvo01}, 
2004S \citep{krisciunas07}, and 2005cf \citep{bufano09}.  
The spectra used for SNe~Ib were from SNe 1998T \citep{matheson01}, 1999dn \citep{deng00}, 
and 2008D \citep{malesani09}. 
For SNe~Ic, spectra were used from SNe 1994I \citep{clocchiatti96}, 1998bw \citep{patat01}, 
2004aw \citep{taubenberger06}, and 2007gr \citep{valenti08}. We included one broad-lined 
SN due to the fact that there are several broad-line SNe~Ic included in this distribution. Any 
variation from this is reflected in the uncertainties.   
The spectra used for SNe~IIb were from SNe 1993J \citep{barbon95} and 2011dh \citep{arcavi11}.  
Due to the lack of publicly available SNe~IIL spectra we used spectra from SNe~IIP.  
The spectra used for both SNe~IIL and IIP were from SNe 1999em \citep{hamuy01}, and 
2006bp \citep{quimby07a}.    
For SNe~IIn, spectra were used from SNe 1995G \citep{pastorello02}, and 1998S \citep{fassia01}. 
The variation in K--correction for each SN was used to estimate the    
uncertainty. The average uncertainties in K--corrections for each SN type are as follows: for SNe~Ia we 
have 0.03 magnitudes; 
for SNe~Ib, 0.10; SNe~Ic, 0.26; SNe~IIb, 0.32; SNe~IIL, 0.10; SNe~IIP, 0.17; SNe~IIn, 0.10. Due to the 
large variation in CCDs (and other detectors), SNe with only unfiltered magnitudes were excluded from 
the study.

\subsection{Distances} 
Distances were handled in the following manner. 
Our first choice was to use cepheid distances. 
If a cepheid distance was known for the host galaxy, then that 
distance was used. If cepheid distances were known for galaxies in the group of galaxies 
containing the host (but no distance known for the host galaxy), then the average of known 
cepheid distances in that group was used for the host galaxy. Cepheid distances were taken from  
\cite{freedman01}, \cite{macri01}, \citet{ferr00,ferr07}, 
\citet{saha01a,saha01b,saha06}, \cite{thim03}, \cite{gieren09}, 
\cite{piet06a,piet06b}, and \cite{soszynski06}. From our entire sample of cepheid distances, 
the average uncertainty is 0.08 magnitudes.   
When cepheid distances were not known, we used distances taken from the
Nearby Galaxies Catalog \citep[NGC;][]{tully88} provided that $cz < 2000$ km s$^{-1}$.  
We adopt an uncertainty of 0.2 magnitudes in the distance modulus for the NGC distances. 
There were a number of nearby SNe for which there was no distance in the NGC or a cepheid distance. 
For nearly all of these, we used distances taken from NED. These distances were primarily determined by 
the Tully-Fisher method.  
There were, however, a few nearby SNe where no distance could be found and 
therefore, those SNe could not be included in this study.    
For SNe beyond $cz = 2000$ km s$^{-1}$, luminosity distances were calculated assuming 
$H_0 = 70\ km\ s^{-1}Mpc^{-1}$, $\Omega_m = 0.3$, and $\Omega_{\Lambda} = 0.7$.   
The distances taken from the NGC were adjusted from H$_0 = 75$ to H$_0 = 70$. 
The overall uncertainty in distance modulus for each type was taken to be the weighted average of the uncertainties  
for the different methods used. The combined uncertainty in distance modulus
for each of the distributions 
is as follows: Ia 0.08 mag., Ib 0.16, Ic 0.14, IIb 0.17, IIL 0.15, IIP 0.19, and IIn 0.14.

\subsection{Extinction} 
Foreground extinction was accounted for according to \cite{schlegel98}. The uncertainties were calculated 
at 16\% of the average extinction for each SN type, as per \cite{schlegel98}. 
For the SN~Ia we have 0.01 mag., for SN~Ib 0.02, SN~Ic 0.04, SN~IIb 0.01, 
SN~IIL 0.03, SN~IIP 0.01, SN~IIn 0.02.  

Host--galaxy extinction, when possible, was estimated from the literature.  
The following sources, and references therein,   
were the primary sources used to help us make these estimates:  
\cite{drout11}, \cite{folatelli10}, \cite{hamuy03}, \cite{harutyunyan08}, \cite{reindl05}, \cite{riess99}, 
\cite{sanders12}, and \cite{shaefer96}. When only the equivalent width of the host \ion{Na}{1} D feature was available, 
we used the relation: A$_V =$ 0.16EW \citep{turatto03} and converted to A$_B$.   
This is the more conservative of the Turatto relations. 

We have noticed that there is a publication bias for host--galaxy extinctions. Those SNe where host--galaxy 
extinction is thought to be significant tend to have the extinction calculated. For those SNe where it is not thought
to be significant, further analysis is not usually carried out. This leads to an average of reported extinctions that is
larger than the expected average. 
Because a host--galaxy extinction estimate was not available for every SN, we used 
a probability distribution for host--galaxy extinction derived by \cite{hatano98}. They determined
the extinction distributions using a Monte--Carlo
technique along with a simple model of the spatial distribution of dust in a typical SN--producing galaxy.
This was done separately for SNe~Ia and for core--collapse SNe.
Due to the fact that highly extinguished SNe are not likely to be discovered, we have truncated the extinction
distributions at 1.5 mag. R02 truncated the distributions at 0.6 mag, but it has become clear that a
significant number of SNe are being discovered with larger host--galaxy extinctions.
The mean model extinction for SNe~Ia is 0.30 magnitudes and the mean for core--collapse SNe is 0.49 magnitudes. 
In order to maintain these mean values, we averaged over all SNe~Ia and all core--collapse SNe
(separately). From this we assigned a value of 0.237 magnitudes to all SNe~Ia   
which do not have a host--galaxy extinction estimate.   
Similarly, we assigned a value of 0.292 magnitudes to all core--collapse SNe without a host--galaxy extinction 
estimate.  
The uncertainties in the model extinction values used here are unknown but thought to be significant. 
We, therefore, estimate the uncertainty in host galaxy extinction to be half the value of the mean model 
extinction (0.15 mag for Type Ia SNe and 0.245 mag for core--collapse SNe).

\section{Analysis} 
The data are displayed here in several ways. Miller diagrams 
are a good way to show not only the data, but also the Malmquist bias.  
In order to account for this selection bias, we used a bootstrap method that 
estimates the data we can't see using the data we can see.  
We also present volume--limited samples, as was done in R02. 
The bias--corrected distributions and the volume--limited distributions are displayed as histograms.

\subsection{Miller Diagrams} 
A Miller diagram is a graph of peak absolute magnitude versus distance modulus, and is therefore a log--log plot. 
Figure~2 is a Miller diagram of all SNe in the ASC for which the peak magnitudes are known. 
For this figure, foreground extinction and host--galaxy extinction have been taken into account. 
The type is not known for every SN in this figure and therefore K--corrections have not been done for this data.  
A few redshifts are given along the top of the graph, corresponding with distance moduli $\mu =$ 35, 40 and 45.  
There are 718 SNe in this figure, which is more than twice the number of maximum--light SNe from R02.  

There are two straight dashed lines in Figure~2 that are found in all of the Miller diagrams presented here. 
The diagonal line represents the apparent magnitude limit used in the bias--correction method discussed in Section~3.2 below.  
The vertical line at $\mu = 35$ shows the limit for 
our volume--limited samples (mentioned at the end of Section~3.2). 
There are two horizontal lines in Figure~2 as well. SNe above the top horizontal line (M$_B = -21$)  
are considered to be superluminous while SNe below the bottom horizontal line (M$_B = -15$) are considered to be subluminous. 
There are similar horizontal lines for the Miller diagrams of the different SN types (Section~4). 
The data in these figures include K--corrections in addition to the foreground and host--galaxy extinction.

\subsection{Malmquist Bias} 
The selection effect of Malmquist bias is obvious in Figure~2, where we see a lack of relatively dim 
SNe at large distances (excluding the high redshift SNe~Ia found in targeted searches). 
In order to correct for this bias, we have developed a bootstrap method that 
estimates, not only the number of SNe, but also the distribution of SNe in incomplete regions of the 
Miller diagram. To do this we use an apparent--magnitude completeness limit (BL). 
This is shown as the diagonal dashed line in Figure~2. 
The portion of the sample with apparent magnitudes brighter 
than this limit (left of the line) are considered to be complete. 
By complete, we don't mean that every SN that actually occurred is accounted for, but rather that we have 
a complete representation of that portion of an unbiased distribution. 
The portion of the sample with apparent magnitudes dimmer than this limit (right of the line) 
are considered to be incomplete. 
From Figure~2, we see that the data used here appears relatively complete for B $<$ 17 (left of the diagonal line).  
We also find that B $=$ 17 is within approximately one standard deviation of the mean apparent magnitude 
for the main bulk of SNe displayed in  
Figure~2 (excluding the high redshift SNe~Ia). Therefore, a magnitude limit of 
BL $=$ 17 is appropriate.   

Next, it is assumed that the true distribution of absolute magnitudes is roughly the same for any relatively large 
volume of space. As a check for this assumption, we looked at two concentric volumes within $\mu = 32$ (25 Mpc), where selection bias 
should be negligible. The only samples that have enough SNe for this check are the SNe~Ia and IIP.  
The first volume is a sphere out to $\mu = 31$ (16 Mpc) and the second is a spherical shell from $\mu = 31$ 
to $\mu = 32$. The mean absolute magnitudes for the two volumes of SNe~Ia are very close, with overlapping 
uncertainties. This is also the case for the SNe~IIP. The standard deviations in the two volumes differ by about 0.2 
magnitudes, for both SNe~Ia and IIP.  
We, therefore, use two volumes in our calculations. The first, volume A, extends out to some  
distance modulus, $\mu_1$. The second, volume B, extends from $\mu_1$ out to a distance modulus of $\mu_2$ 
(a concentric spherical shell). 
An absolute--magnitude limit, ML, is chosen such that the region of volume B that is brighter than this 
limit is considered to be complete, while the dimmer region is incomplete. Both the dim and bright regions of 
volume A are considered to be complete. ML, $\mu_1$, and $\mu_2$ are determined by BL.  
We now have four regions as shown in Figure~3a: 
A$_{bright}$, A$_{dim}$, B$_{bright}$, and B$_{dim}$. 

If all volumes have roughly the same distribution, then the ratio of dim to bright SNe in one volume should 
be the same as for another volume for some brightness limit, ML.   
Therefore, the ratio of the number of SNe in region A$_{dim}$ to the number in region A$_{bright}$ must be equal to the ratio of the 
number in region B$_{dim}$ to the number in region B$_{bright}$. From this we can find the number of SNe 
missing from region B$_{dim}$. The distribution of these new SNe in region B$_{dim}$ is determined by the 
distribution of SNe in region A$_{dim}$ (which is considered to be a complete region).  
The absolute magnitudes and distance moduli of the new SNe are generated randomly within the appropriate 
one--magnitude by one--magnitude cell. The absolute--magnitude limit (ML) and the distance modulus limits  
($\mu_1$ and $\mu_2$) are incremented by intervals of one magnitude. When there are no more SNe in one of  
the two bright regions (considered to be complete), 
the process is stopped and the distribution is truncated at the last $\mu_2$. 
The method starts by finding the dimmest SN in the sample that has an apparent magnitude brighter than BL.  
We set the limiting absolute magnitude value, ML, above that point. Then $\mu_1$ and $\mu_2$ are found according to ML and BL.   
The panels in Figure~3 show the steps of the process for the SN~Ic distribution. The open circles represent the original data, while the 
the filled circles represent the new SNe added to account for the selection bias. 
In the first step (Figure~3a), there are no new SNe added to the distribution. In the second step (Figure~3b), 
the limits are incremented and several SNe are added to the region B$_{dim}$. More are added in step three (Figure~3c)
after the limits are incremented once more. 
In the next step (Figure~3d), as well as any later steps, there are no more SNe in B$_{bright}$.  
Therefore, the resulting distribution does not extend beyond $\mu = 36$, which is the last point the process is able 
to make an estimate of the missing SNe.  

These bias--corrected distributions were calculated ten times for each SN type. While there was some variation, 
due to the random generation of new data within a cell, that variation was small. Uncertainty due to this variation 
ranges from 0.004 magnitudes for SNe~Ia to 0.056 magnitudes for SNe~Ib. The variation in standard deviation was similar.  

Volume--limited samples from the original data were also used for comparison. We limited these samples to radial distances of  
$\mu < 35$ (100 Mpc). The volume--limited samples used in R02 were limited to $\mu < 40$ (1 Gpc), however 
because we now have significantly more SNe to work with, we are able to use a smaller volume.

\section{Results} 
Table~1 shows the results for the bias--corrected samples and Table~2 shows the results for the 
volume--limited samples. Both tables give the mean absolute magnitudes where foreground extinction, 
host--galaxy extinction and K--corrections have been taken into account.  
The uncertainties given are the total uncertainties with contributions from apparent magnitude, distance modulus, 
foreground and host--galaxy extinction as well as for K--corrections. The uncertainties in Table~1  
also include a contribution from the uncertainties for the bias--correction process. 
Superluminous and subluminous SNe will be discussed below for the entire sample and for the separate SN types. 

In order to compare the results here with those of R02, one must keep in mind that the distributions in
R02 were volume--limited samples limited at $\mu = 40$ rather than $\mu = 35$.  
Also, host--galaxy extinction
in R02 was handled statistically for each distribution as a whole without consideration for the 
extinction of individual SNe.
The bias--correction method used in the current study depends heavily on the dimmest SNe and it is
therefore very important to know the host--galaxy extinction as well as possible for those dim SNe.
It is for this reason that we considered host--galaxy extinction for as many SNe as possible, especially
for the dimmest SNe. 
For the sake of comparison, we have also calculated distributions for each of the SN types without the 
host--galaxy extinction removed. These comparisons are given for each of the SN types in the subsections 
below ($4.3-4.9$).  

A study by \citet[][hereafter L11]{li11} presented findings of the Lick Observatory Supernova Search (LOSS).   
Data was collected through a controlled search with 
consistent data reduction for 175 SNe in volume--limited samples. They presented absolute R magnitude 
distributions for the various SN types without the removal of host--galaxy extinction. 
They accounted for completeness (selection bias) in their data by using the 
control time for each SN in their volume--limited samples (information we don't have for most of our SNe). 
In some cases, our results are similar to theirs, however some differences are significant.  
Those differences are discussed below (Subsections $4.3-4.9$).

\subsection{Superluminous Supernovae}
In this study, we have nine SNe that are superluminous (M$_B < -21$).
The superluminous SNe shown in Figure~2 are 2003ma (IIn), 2005ap (IIn), 2006gy (IIn), 2006oz (Ib),
2008am (IIn), 2008es (IIL), 2008fz (IIn), 2009jh (IIn), and 2010gx (Ic).
Notice that most of these are SNe~IIn. All of these SNe were discovered after R02. At that time
there were 20 SNe for which M$_B < -20$. In the current study, there are 50.

In an attempt to find the true fraction of superluminous supernovae, we used the method
mentioned above to produce a bias--corrected sample of all SNe for which peak magnitudes are known.   
Our bias--corrected sample extended out to $\mu = 40$ and included 2025 SNe.  
Out of these, only three were superluminous. We therefore consider
approximately $0.1\%$ of all SNe to be superluminous. 
There exist a few superluminous SNe, without IAU designations, that were not included here \citep{quimby13}.  
When taking these SNe into account, the results don't significantly change. This is because nearly all of them 
are at distances greater than $\mu = 40$, where our bias--corrected sample stops. Therefore, these SNe 
exist in a much larger volume that cannot be corrected for selection bias and it is reasonable that they make up 
a very small fraction of SNe in that large volume.  

When considering SNe for which M$_B < -20$,
we find that there are 84 out of 2025 in our bias--corrected sample. We therefore find that approximately
$4\%$ of all SNe are brighter than M$_B = -20$.

\subsection{Subluminous Supernovae}  
There are seven SNe in Figure~2 that we consider to be subluminous (M$_B > -15$).   
SNe 1923A (possibly a IIP) and 1945B (unknown type) were both found to be far from the 
center of a nearly face--on galaxy (M83) and therefore are not thought to be heavily extinguished 
\citep{shaefer96}. Because of this we estimated their host--galaxy extinction to be zero. 
This extinction is highly uncertain. If the average model extinction for core--collapse SNe were to be applied to  
these SNe, 1923A would still be subluminous, but 1945B would not. 
It is unusual to find a Type~Ia in this category, but SN~2008ha appears to be intrinsically dim.  
As a more extreme version of SN~2002cx \citep{foley09,foley10} it is definitely peculiar.  
There is the possibility that it is not a Type Ia SN at all, but rather a core--collapse SN \citep{valenti09}. 
The host galaxy of SN~1940E is dusty and nearly edge--on \citep{shaefer96}. Because of this, 
it is thought to have significant extinction from its host galaxy. This extinction is very uncertain 
and we have estimated it to be one magnitude. It could be somewhat larger than this, but it is 
still likely to be subluminous.  
Due to the lack of narrow interstellar \ion{Na}{1} D lines in the spectrum of SN~1999br \citep{pastorello04}, 
we have estimated its host--galaxy extinction to be zero. If the average model host--galaxy extinction 
were to be applied to this SN, it would still be subluminous. The host--galaxy extinction for SN~2003bk is 
not known. If the average model extinction were applied, it would not be subluminous.     

We used the same bias--corrected sample mentioned above for superluminous SNe in order to find the percent 
of all SNe that are subluminous. 
We found that approximately $3\%$ of all SNe have peak absolute magnitudes for which M$_B > -15$.

\subsection{Type Ia} 
Figure~4 is a Miller diagram of all SNe~Ia not known to be spectroscopically peculiar. 
Because a significant source of our data is the ASC, peculiarity is not known for all of our SNe~Ia. 
Therefore, our sample may include a few peculiar SNe~Ia. 
As with the Miller diagrams for the other types, foreground extinction, host--galaxy extinction and 
K--corrections have been taken into account. There are 382 SNe~Ia in this graph.  
Figure~5 shows two absolute--magnitude distributions determined from the sample of SNe~Ia shown in Figure~4. 
The top panel is a histogram of the bias--corrected sample while the bottom panel is a histogram of the 
volume--limited sample. The data has been collected into bins with a width of one magnitude.   
The bias--corrected sample has a mean absolute magnitude of M$_B = -19.25 \pm 0.20$ and a standard
deviation of 0.50 for 239 SNe (Table~1). 
The bias--correction process for SNe~Ia continued out to $\mu = 38$, and stopped at that point.   
The results for the volume--limited sample ($\mu < 35$) are virtually the same. The mean absolute magnitude is 
M$_B = -19.26 \pm 0.20$ and the standard deviation is 0.51 for 171 SNe (Table~2).  
Selection bias clearly does not play a significant role in the normal SN~Ia distribution. 

There are three horizontal lines in Figure~4. The center line represents the mean of the bias--corrected 
distribution ($-$19.25). 
The other two lines represent the $2\sigma$ limits ($-$18.25 and $-$20.25). Magnitudes outside these limits 
are considered extreme for this type of SN.  
Similar lines are shown on the Miller diagrams for the other types as well.   
SN~1999fv is the brightest SN~Ia in the study. Even though its 
redshift is somewhat uncertain \citep{coil00}, it still appears to be a very bright Ia.  
SN~2005W suffers from about 1.4 magnitudes of host--galaxy extinction \citep{folatelli10} and with this 
taken into account it is barely brighter than the $2\sigma$ limit.  
Little has been reported for SN~2004if, so it is not clear if it is actually as bright as it appears.  
There are seven SN~Ia that fall below the $2\sigma$ line. 
For four of these (1963P, 1997O, 2004hu, and 2004hw), not much is reported in the literature.  
It is not clear if they are intrinsically dim SNe~Ia. 
SN~1998dm suffers from significant host--galaxy extinction. This  
extinction is not well known and if our conservative estimate is an under estimate, then it might not actually  
be subluminous. SN~2002jg suffers from a minimal amount of host--galaxy extinction. This extinction is not well known 
and if the actual extinction is only slightly higher than our estimate, then this SN would not be considered subluminous.  
However, SN~1996ai does appear to be intrinsically dim, even when accounting for a significant amount of host--galaxy extinction.  

We find that our distribution with host--galaxy extinction taken into account is 0.51 magnitudes brighter than our distribution 
where host--galaxy extinction has not been taken into account. This is different from the mean model extinction, 
due in part because the  
mean model extinction was averaged over all SNe~Ia in the study, most of which are beyond $\mu = 37$ where the 
bias--correction process stops.  
However, another reason for this difference is that there is a large number of dim SNe~Ia that suffer a 
considerable amount of host--galaxy extinction. This not only makes the overall distribution brighter, but also 
makes the standard deviation much smaller. The standard deviation for the distribution where host--galaxy extinction 
is not taken into account is 0.98, which is almost twice the that of the distribution where host--galaxy extinction is 
taken into account. 

Compared to R02, the current sample of SNe~Ia is 0.12 magnitudes brighter (after adjusting for a different H$_0$). 
When comparing our results to those of L11, we find that their mean absolute magnitude is 
0.49 magnitudes dimmer than ours (after adjusting for a different filter and a different H$_0$). When not considering the  
effects of host--galaxy extinction, the difference reduces to 0.02 magnitudes.

\subsection{Type Ib} 
R02 combined SNe Ib and Ic due to their small numbers. In this study, the numbers are large 
enough that we can treat them separately. Figure~6 is a Miller diagram of the SN~Ib sample.  
There are 20 SNe shown in this figure. The absolute--magnitude distributions for SNe~Ib are shown in Figure~7.  
The mean absolute peak magnitude for the bias--corrected distribution is $-17.45 \pm 0.33$ 
with a standard deviation of 1.12 magnitudes for 38 SNe. This puts the two--sigma limits in Figure~6 at $-$15.21 and $-$19.69.    
The volume--limited sample of 18 SNe has a mean absolute peak magnitude of 
M$_B = -17.54 \pm 0.33$ and a standard deviation of 0.94. This is roughly the same as the bias--corrected distribution.  

The brightest SN in Figure~6 is SN~2006oz. It is thought to have reached  
its high luminosity through the strong interaction between the SN ejecta and dense circumstellar material 
\citep{moriya12}. The other extremely bright Ib is SN~1991D. It had a relatively low photospheric velocity 
for such a bright SN and does not fit conventional light--curve models \citep{benetti02,richardson06}.   
\cite{benetti02} suggested that it might be a SN~Ia exploding inside the extended He envelope of a 
companion star. 
There was not enough information in the original sample to make any determination about
the selection bias beyond $\mu = 37$. Because of this, the brightest SN (2006oz) was not included in the bias--corrected 
sample. Superluminous SNe like SN2006oz are likely to make up only a very small   
fraction of all SNe~Ib. There are no extremely dim SNe~Ib.  

When host--galaxy extinction is not considered, then the bias--corrected distribution is dimmer by 0.85 mag 
and the volume--limited distribution is dimmer by 0.87 magnitudes. 
Our mean absolute magnitude is 1.18 mag brighter than that reported in L11. When not accounting for host--galaxy 
extinction, the difference reduces to 0.33 magnitudes.

\subsection{Type Ic} 
A Miller diagram of all SNe~Ic is shown in Figure~8. There are 49 SNe in this figure.  
Figure~9 shows the absolute--magnitude distributions for SNe~Ic. 
The bias--corrected sample has a mean absolute--magnitude of M$_B = -17.66 \pm 0.40$ and a standard deviation 
of 1.18 for 53 SNe. Therefore, the $2\sigma$ lines in Figure~8 are $-$15.30 and $-$20.02.  
The bias--correction process stopped at $\mu = 36$. 
The volume--limited sample has a mean absolute--magnitude of M$_B = -17.67 \pm 0.40$ and a standard deviation 
of 1.04 for 36 SNe. The bias--corrected distribution is virtually the same as the volume--limited distribution.  

There are four SNe in Figure~8 considered to be extremely bright for SNe~Ic. They are SNe 1999cq, 2007D, 2007bi and 2010gx.     
SN~2010gx is far brighter than the others. \cite{pastorello10} show SN~2010gx to have a broad--lined spectrum.  
Its peak luminosity is likely to be powered by interaction with its circumstellar wind \citep{ginzburg12}.  
SN~1999cq is a peculiar, and possibly unique, SN~Ic \citep{matheson00}. 
SN~2007D is a broad--lined SN~Ic with a considerable amount of host--galaxy extinction \citep{drout11}, 
which was accounted for here. 
SN~2007bi is thought to be the pair-instability explosion of a star with a core mass of $\sim100 M_{\odot}$ \citep{galyam09}. 
An alternative explanation for its extreme brightness is that a much lower--mass star collapses to form a magnetar \citep{kasen10}.  
The brightness of the SN is a result of the energy gained from the spin-down of this newly--formed magnetar.  
SN~2006eg is the only SN in Figure~8 that is considered to be extremely dim. 
The host--galaxy extinction is not known for this SN and it is therefore not clear if  
it is actually subluminous for a SN~Ic.  

R02 had considered the possibility that the combined sample of SNe~Ibc might be separated into two groups (bright and normal) 
and concluded that there was not enough data at that time to determine if this separation was warranted.  
It is clear from Figures~7 \& 9 that this separation no longer exists.  

The bias--corrected distribution is brighter by 1.31 mag than it would be if host--galaxy extinction were not taken into account.  
Similarly, the volume--limited distribution is brighter by 0.92 magnitudes. This is considerably more than the mean model extinction 
for core--collapse SNe (as was the case for the SNe~Ib as well). However, as we will see below, this is compensated 
for by the fact that the Type II SNe have noticeably less host--galaxy extinction than the model average.  
Our mean is 1.74 magnitudes brighter than that of L11. When using our distribution without the effects of 
host--galaxy extinction taken into account, the difference is reduced to 0.43 magnitudes.

\subsection{Type IIb} 
SNe~IIb were not included in R02 due to the lack of data at that time. While the lack of data is still an issue, 
our sample is large enough for a reasonable comparison with other types. Figure~10 is a Miller diagram of 15 SNe~IIb.  
The absolute--magnitude distributions for SNe~IIb are shown in Figure~11. 
The bias--corrected sample has a mean absolute--magnitude of M$_B = -16.99 \pm 0.45$ and a standard deviation of 0.92 
for 16 SNe. Only one SN was added to this distribution and due to the fact that there are no original SN~IIb 
beyond $\mu = 35$, the bias--correction process stops at that point.  
The volume--limited sample is roughly the same with a mean absolute--magnitude of M$_B = -17.03 \pm 0.45$ and a 
standard deviation of 0.93 for 15 SNe.  
The $2\sigma$ limits for the bias--corrected distribution are $-$15.15 and $-$18.83 and we see that there are 
no extreme SNe~IIb in Figure~10. A few key SNe are labeled in Figure~10.  

The bias--corrected distribution is brighter by 0.41 mag than it would be if host--galaxy extinction were not taken 
into account. The volume--limited distribution is brighter by 0.37 magnitudes.   
Our mean is 0.67 mag brighter than that of L11 and reduces to 0.26 mag after removing the  
effects of host--galaxy extinction.

\subsection{Type IIL} 
There are 19 SNe~IIL plotted in the Miller diagram shown in Figure~12. 
The absolute--magnitude distributions for SNe~IIL are shown in Figure~13.
The bias--corrected sample has a mean absolute magnitude of M$_B = -17.98 \pm 0.34$ and a standard deviation of 0.86 
for 19 SNe. This distribution was truncated at $\mu = 36$ and therefore did not include the brightest SN in Figure~12
(2008es). Only one SN was added in the process. Therefore, as expected, the volume--limited sample is very similar 
with the same mean absolute magnitude, M$_B = -17.98 \pm 0.34$, and a slightly different standard deviation, 0.90, for 17 SNe.   

There are two SNe in Figure~12 considered to be extremely bright SNe~IIL (1979C and 2008es).  
The brightness of SN~1979C may be due to either a large ejected mass or interaction of the ejecta with the 
circumstellar material of the progenitor star, or both \citep{fesen93}.  
The extreme luminosity of SN~2008es is possibly due to both, a very large ejected mass and the strong interaction 
of the ejecta with a dense circumstellar material \citep{miller09}.  
However, this explanation may not be sufficient \citep{gezari09}. \cite{kasen10} suggest that the 
rotational energy from a remnant magnetar is providing the extra energy. 
There are no subluminous SNe~IIL.  

Both the bias--corrected and volume--limited distributions with host--galaxy extinction considered are brighter 
by 0.36 magnitudes than when host--galaxy extinction is not considered.   
When compared to R02, the current sample is 0.28 magnitudes brighter. 
Our mean is 0.42 magnitudes brighter than that of L11. If the effects of host--galaxy 
extinction are not taken into account, the difference is reduced to 0.06 magnitudes.

\subsection{Type IIP} 
The Miller diagram for 74 SNe~IIP is shown in Figure~14. 
The absolute--magnitude distributions for SNe~IIP are shown in Figure~15.
The bias--corrected sample has a mean absolute--magnitude of M$_B = -16.75 \pm 0.37$ and a standard deviation of 0.98
for 78 SNe. This is the dimmest mean absolute magnitude of any of the SN types.  
Only four SNe were added in the bias--correction process and none of the original data were found beyond $\mu = 35$. 
The bias--corrected sample is therefore very similar to the volume--limited sample.  
The volume--limited sample has a mean absolute--magnitude of M$_B = -16.80 \pm 0.37$ and a standard deviation of 0.97.  
This sample contains all of the original 74 SNe~IIP.

If the approximately 5 magnitudes of host--galaxy extinction \citep{meikle02,pozzo06} were not taken into account, 
SN~2002hh would appear to be extremely dim. Instead, it is barely brighter than the $2\sigma$ limit of M$_B = -18.71$.   
It is the only SN considered to be extremely bright for SNe~IIP.  
SN~1999br is the only SN~IIP that is dimmer than the lower $2\sigma$ limit of M$_B = -14.79$.  
It appears to not have any significant host--galaxy extinction \citep{pastorello04} and is therefore extremely 
dim even for a Type IIP SN. 

When accounting for host--galaxy extinction, our bias--corrected mean is 0.46 mag brighter than it would be if we 
did not account for host--galaxy extinction. Similarly, our volume--limited distribution is brighter by 0.45 magnitudes.  
The results of the current sample are roughly the same as that reported in R02. 
Our mean is 0.96 magnitudes brighter than that of L11.  
When removing the effects of host--galaxy extinction, this reduces to 0.50 magnitudes. While this is considerable, 
it still lies within the overlapping error bars of the two distributions.

\subsection{Type IIn} 
The Miller diagram for 29 SNe~IIn is shown in Figure~16. 
The absolute--magnitude distributions for SNe~IIn are shown in Figure~17.
The bias--corrected sample has a mean absolute--magnitude of M$_B = -18.53 \pm 0.32$ and a standard deviation 
of 1.36 for 48 SNe. 
The volume--limited sample has a mean absolute--magnitude of M$_B = -18.62 \pm 0.32$ and a standard deviation 
of 1.48 for 21 SNe. This distribution is only 0.09 magnitudes brighter than the bias--corrected distribution.  
The bias--correction process extended out to $\mu = 39$ which allowed 
for more dim SNe to be generated while going out far enough to include only two superluminous SNe.  
Apparently, these superluminous SNe only make up about 4\% of all SNe~IIn, 
which is still considerably larger than for all SNe in general 
(approximately 0.1\%, see section~4.1 above).  

The $2\sigma$ limits are $-$15.81 and $-$21.25. This is the widest distribution in the study.  
Even with the brightest upper limit of any other SN type, there are still four SNe~IIn brighter than 
this limit. They are 2005ap, 2006gy, 2008am, and 2008fz. Two other SNe (2003ma and 2009jh) considered to be generally 
superluminous above, fall just below the upper $2\sigma$ limit here and therefore are not considered extremely  
bright for SNe~IIn.  
Notice that after the K--correction is applied, SN~2003ma actually falls below M$_B = -21$.  
SN~2006bv is the only subluminous SN~IIn. It appears not to have had a large 
host--galaxy extinction \citep{smith11} and was therefore intrinsically subluminous. However, there is 
the possibility that it was an eruption of a luminous blue variable and not a SN at all \citep{smith11}.  

The bias--corrected distribution is 0.42 mag brighter than it would be if the effects of host--galaxy extinction 
were not taken into account. The volume--limited distribution would be 0.37 mag brighter. 
The current sample is 0.29 magnitudes dimmer than that reported in R02.  
Compared to the LOSS sample, our sample is significantly larger and, apart from the extremely bright SNe~IIn, 
has more generally bright SNe~IIn.  
The result is a mean that is 1.70 magnitudes brighter than that of L11. After removing the effects of host--galaxy 
extinction, the difference is still considerable at 1.28 magnitudes.

\section{Summary} 
In this study, we collected basic data for as many SNe as possible, mainly from the Asiago Supernova Catalogue. 
From that basic data, we calculated absolute magnitudes for each SN in a self--consistent manner.  
Most of the apparent magnitudes were reported in the B filter. If B--filter peak magnitudes were not available, 
we approximated the K--corrections using spectra from SNe of the same type. We also estimated the 
host--galaxy extinctions for as many individual SNe as possible and used a model to assign values for the rest.  
In order to account for Malmquist bias, we used a bootstrap method to generate bias--corrected distributions 
for seven SN types. These distributions were then compared to volume--limited distributions.  

We found that the distribution with the brightest peak mean absolute magnitude (Ia) was 0.72 magnitudes 
brighter than the next brightest (IIn). 
The SNe Ib and Ic have similar mean absolute magnitudes (M$_B = -$17.45, $-$17.66 respectively). 
Their average is roughly 1.7 magnitudes dimmer than that of the SNe~Ia. 
Type IIb SNe were roughly half a magnitude dimmer than the other stripped--envelope SNe (Ib \& Ic).  
The dimmest type in the study is the SN~IIP distribution which is 1.23 magnitudes
dimmer than the SNe~IIL distribution.
The mean absolute magnitude for the dimmest distribution (IIP) is 2.5 magnitudes dimmer than that
for the the brightest (Ia).

We found that volume--limited samples (limited to $\mu < 35$) produce good approximations of the unbiased
absolute--magnitude distributions for all SN types. 
For most of the distributions, the bias--correction process didn't go far beyond $\mu = 35$ and so we would expect 
them to be similar. However, the SN~Ia and IIn samples went out to $\mu = 38$ and 39 respectively.  
The SN~Ia distributions were nearly identical and the SN~IIn distributions differed by only 0.09 mag in mean and 
by only 0.12 in standard deviation. 
As expected, the SNe~Ia have the smallest standard deviation and are highly concentrated within the $2\sigma$ limits. 
This distribution also has the smallest fraction of extreme SNe (2.6\%), not counting the small sample of SN~IIb with 0\%. 

When comparing the results of the current distributions to those of R02, we found that the two biggest 
differences were for the SNe~IIL and IIn distributions.  
The large number of superluminous SNe~IIn discovered during 
the last decade played a role in this difference. We should also note that the SN~Ia distribution is 
somewhat brighter than that of R02 and this is most likely due to the fact that most of the dimmest SNe~Ia suffered from 
significant host--galaxy extinction. Once that was taken into account, the mean absolute--magnitude became 
brighter.    

We also compared our results to those of the Lick Observatory Supernova Search (LOSS) as 
reported in L11. After adjusting their values for a different H$_0$, different filter (R to B) and making an adjustment 
for host--galaxy extinction, we found that most of their distributions are relatively  
close to ours (within the, sometimes considerable, error bars of the two distributions). 
However, there is one noticeable exception, the SN IIn distribution. As mentioned above, our sample is considerably 
larger (containing all of the LOSS sample) and simply has more bright SNe~IIn than the LOSS sample alone.  

There are nine superluminous SNe (M$_B < -21$) when considering all types. All of these were discovered 
during the last decade and six of them are SNe~IIn.  
From the bias--corrected sample representing SNe of any type, we see that superluminous SNe 
make up only about 0.1\% of all SNe.  
There are seven SNe in this study that are considered to be subluminous 
(M$_B > -15$). We see that approximately 3\% of all SNe in the 
bias--corrected sample are subluminous. 

We would like to thank David Branch for his helpful suggestions. 
This work was partially supported by the National Science Foundation under Grant
No. CHE-1005026. 
This work was also supported by the U.S. Nuclear Regulatory Commission under Grant No. 23N811. 
This research has made use of the NASA/IPAC Extragalactic Database (NED) which is operated by the 
Jet Propulsion Laboratory, California Institute of Technology, under contract with the National 
Aeronautics and Space Administration.

{}

\pagebreak 

\begin{deluxetable}{lccccc}  
\tabletypesize{\scriptsize} 
\tablecaption{Bias--Corrected Distributions \label{table1}}
\tablewidth{0pt} 
\tablehead{ 
\colhead{SN Type} &
\colhead{$\overline{M}_{B}^a$} &
\colhead{$\sigma^b$} & 
\colhead{N} &  
\colhead{$\mu_{limit}$}   
}
\startdata
Ia         & -19.25 $\pm$ 0.20 & 0.50 & 239 & 38 \\  
Ib         & -17.45 $\pm$ 0.33 & 1.12 & 38 & 37 \\  
Ic         & -17.66 $\pm$ 0.40 & 1.18 & 53 & 36 \\  
IIb        & -16.99 $\pm$ 0.45 & 0.92 & 16 & 35 \\  
IIL        & -17.98 $\pm$ 0.34 & 0.86 & 19 & 36 \\  
IIP        & -16.75 $\pm$ 0.37 & 0.98 & 78 & 35 \\  
IIn        & -18.53 $\pm$ 0.32 & 1.36 & 48 & 39 \\  
\enddata 
\tablenotetext{a} {Includes uncertainties in apparent magnitude, distance modulus, foreground and host--galaxy extinction
as well as for K--corrections and the bias--correction process added in quadrature.} 
\tablenotetext{b} {This is the statistical standard deviation in the mean.}

\end{deluxetable} 

\begin{deluxetable}{lccccc}
\tabletypesize{\scriptsize}
\tablecaption{Volume-Limit Distributions \label{table2}} 
\tablewidth{0pt}
\tablehead{
\colhead{SN Type} &
\colhead{$\overline{M}_{B}^a$} &
\colhead{$\sigma^b$} &
\colhead{N}
}
\startdata
Ia         & -19.26 $\pm$ 0.20 & 0.51 & 171 \\
Ib         & -17.54 $\pm$ 0.33 & 0.94 & 18 \\
Ic         & -17.67 $\pm$ 0.40 & 1.04 & 36 \\
IIb        & -17.03 $\pm$ 0.45 & 0.93 & 15 \\
IIL        & -17.98 $\pm$ 0.34 & 0.90 & 17 \\
IIP        & -16.80 $\pm$ 0.37 & 0.97 & 74 \\
IIn        & -18.62 $\pm$ 0.32 & 1.48 & 21 \\
\enddata
\tablenotetext{a} {Includes uncertainties in apparent magnitude, distance modulus, foreground and host--galaxy extinction
as well as for K--corrections added in quadrature.} 
\tablenotetext{b} {This is the statistical standard deviation in the mean.}  

\end{deluxetable} 

\begin{figure}
\plotone{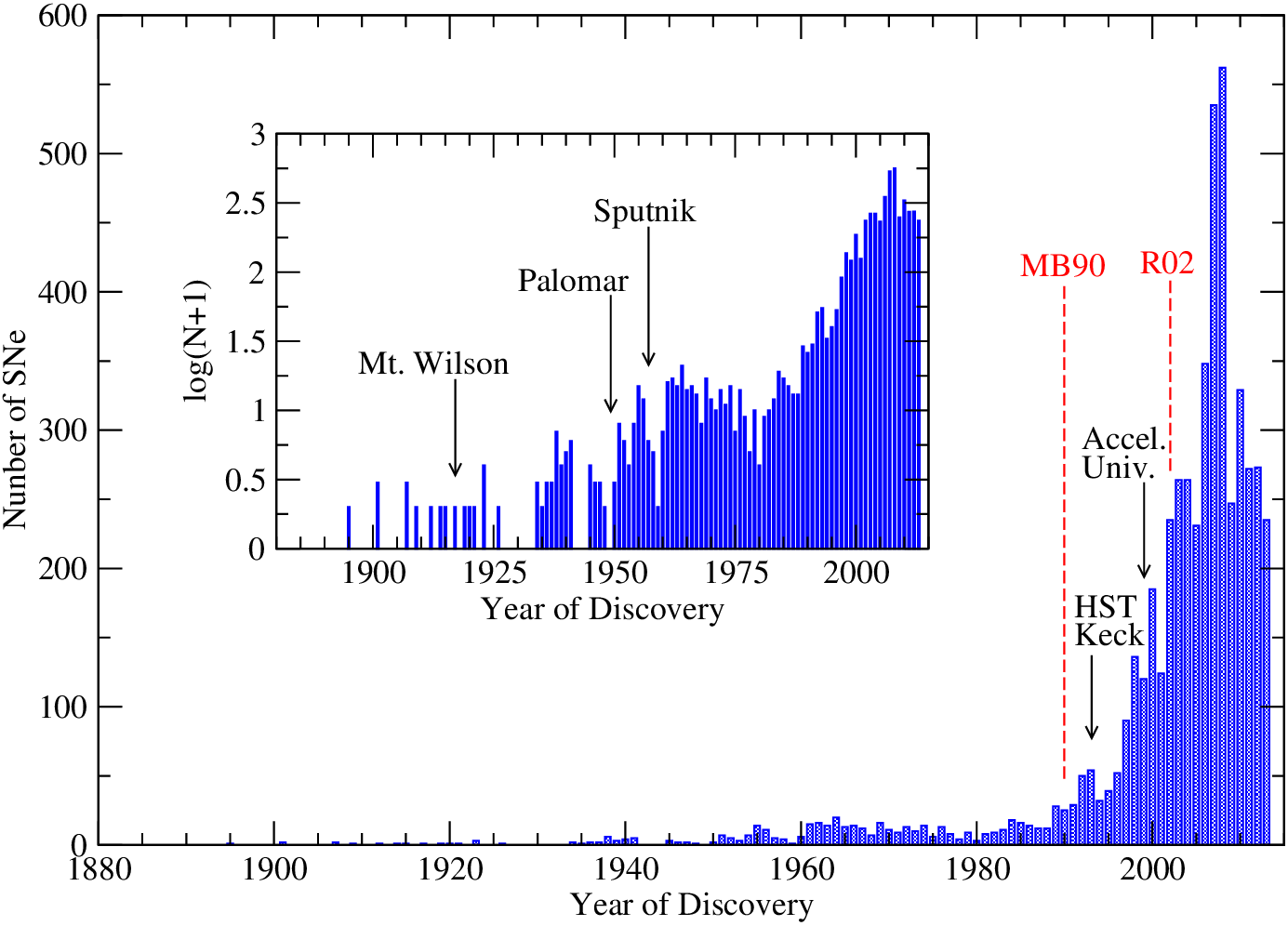}
\caption{\label{fig1}
This is a histogram showing the number of SNe discovered each year as given by the Asiago Supernova Catalogue.  
The inset shows the same data on a log scale.  
}
\end{figure}  

\begin{figure}
\plotone{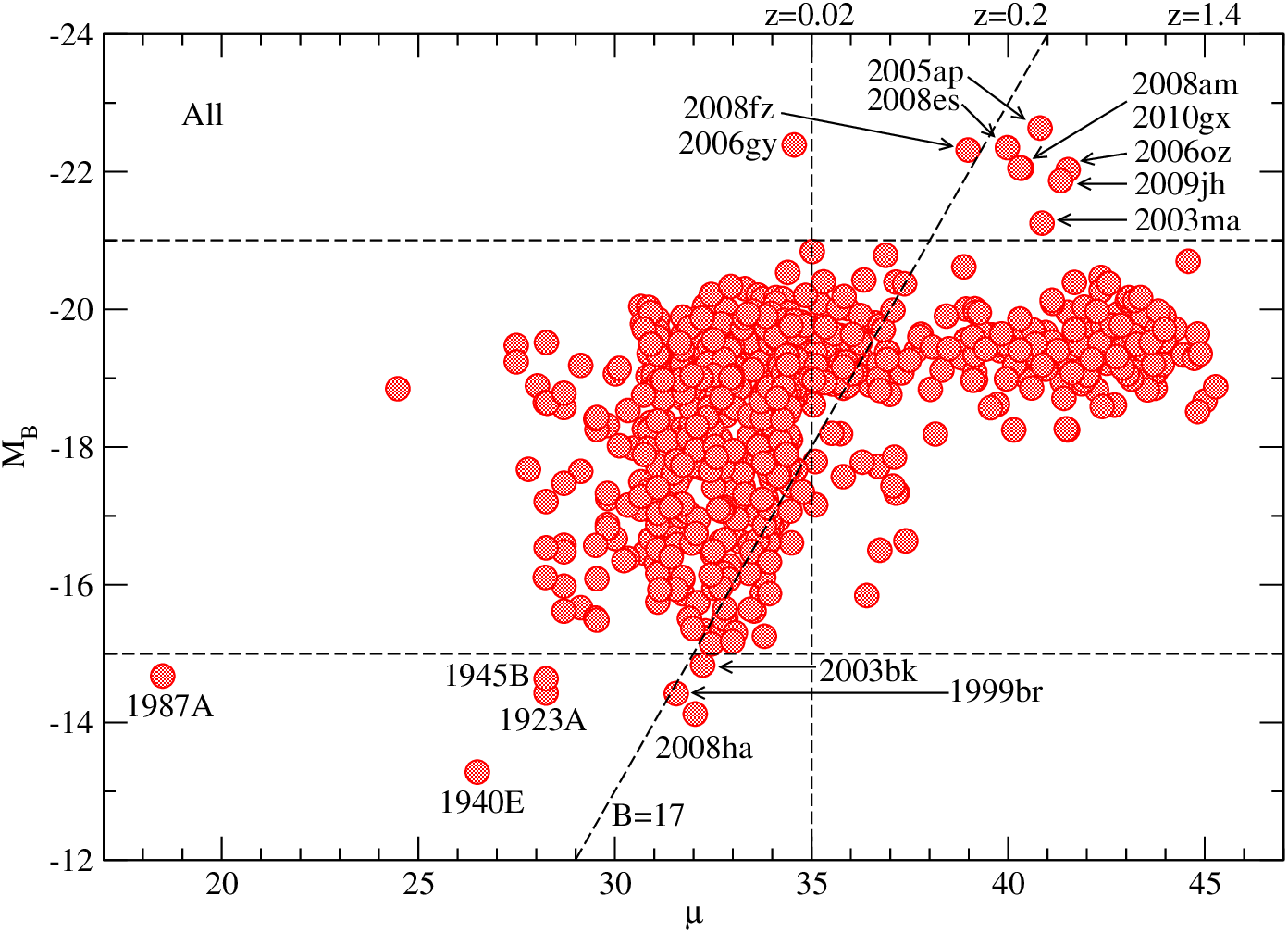}
\caption{\label{fig2}
The peak absolute magnitudes of all 718 supernovae for which peak magnitudes are available 
are plotted here against distance modulus (Miller diagram).  
}
\end{figure} 

\begin{figure}
\plotone{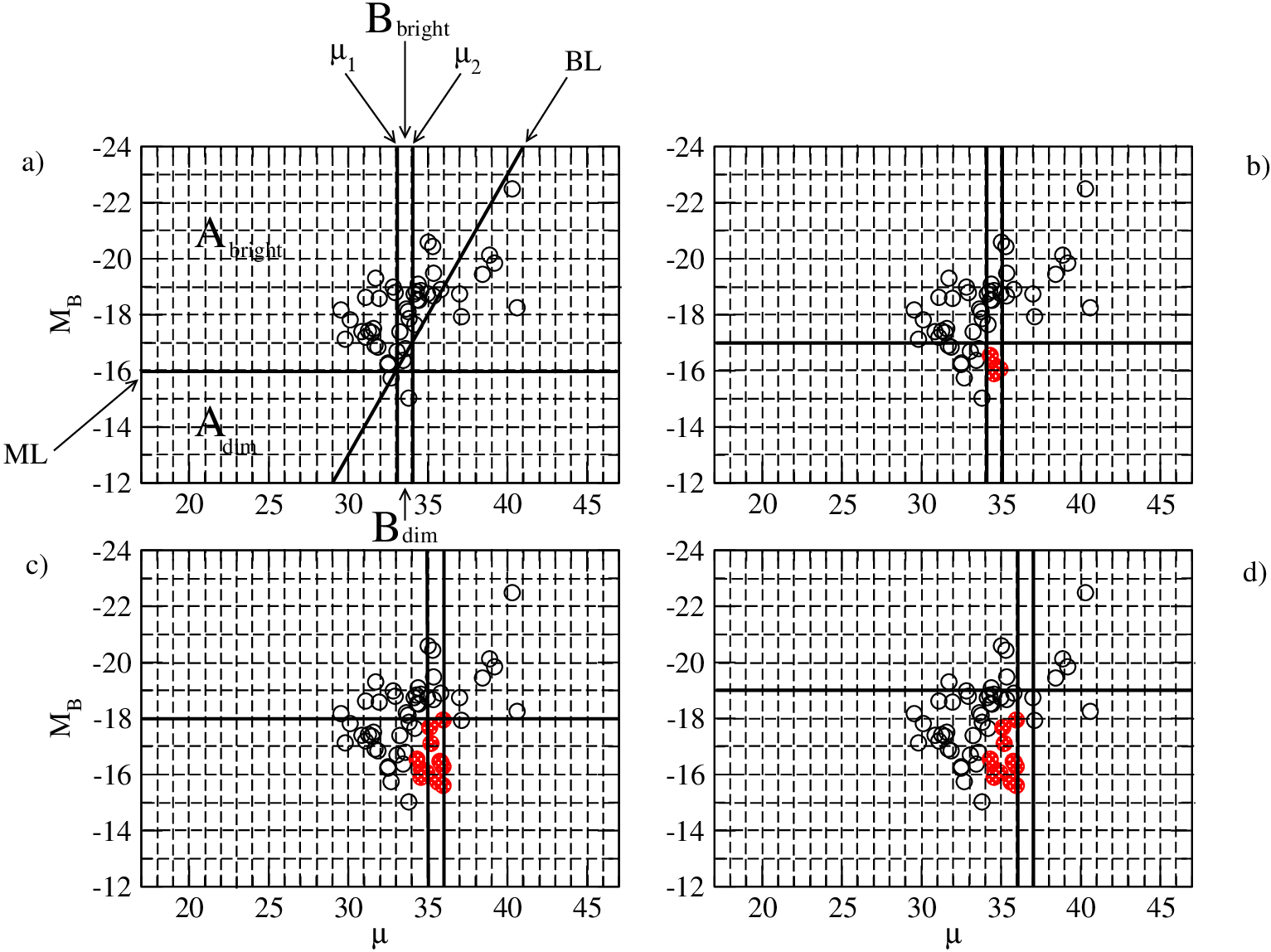}
\caption{\label{fig3}
Displayed here are Miller diagrams for the SN~Ic sample over several steps of the bias--correction process. 
In the first figure (a), we see volume A to the left of $\mu_1 = 33$ and volume B between $\mu_1$ and  
$\mu_2 = 34$. Bright and dim SNe are separated at ML = $-$16. In these figures, open circles are the original data while 
filled circles are SNe that were added to account for selection bias. In each subsequent figure, we see a 
step where the limits have been incremented and data have been added to account for Malmquist bias. 
The process stops at $\mu = 36$.  
}
\end{figure} 

\clearpage 

\begin{figure}
\plotone{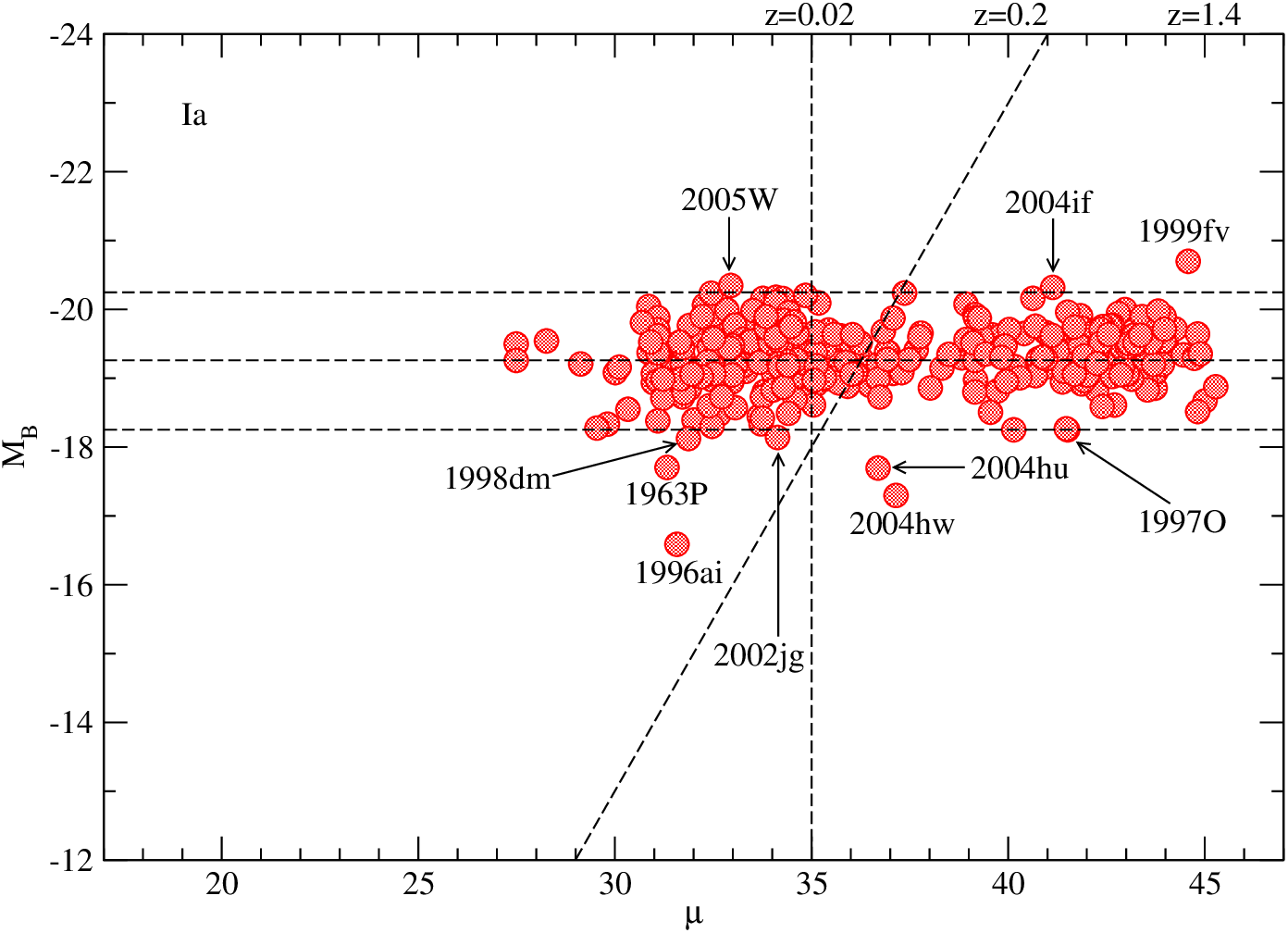} 
\caption{\label{fig4}
Miller diagram of 382 normal SNe~Ia. 
}
\end{figure} 

\begin{figure}
\plotone{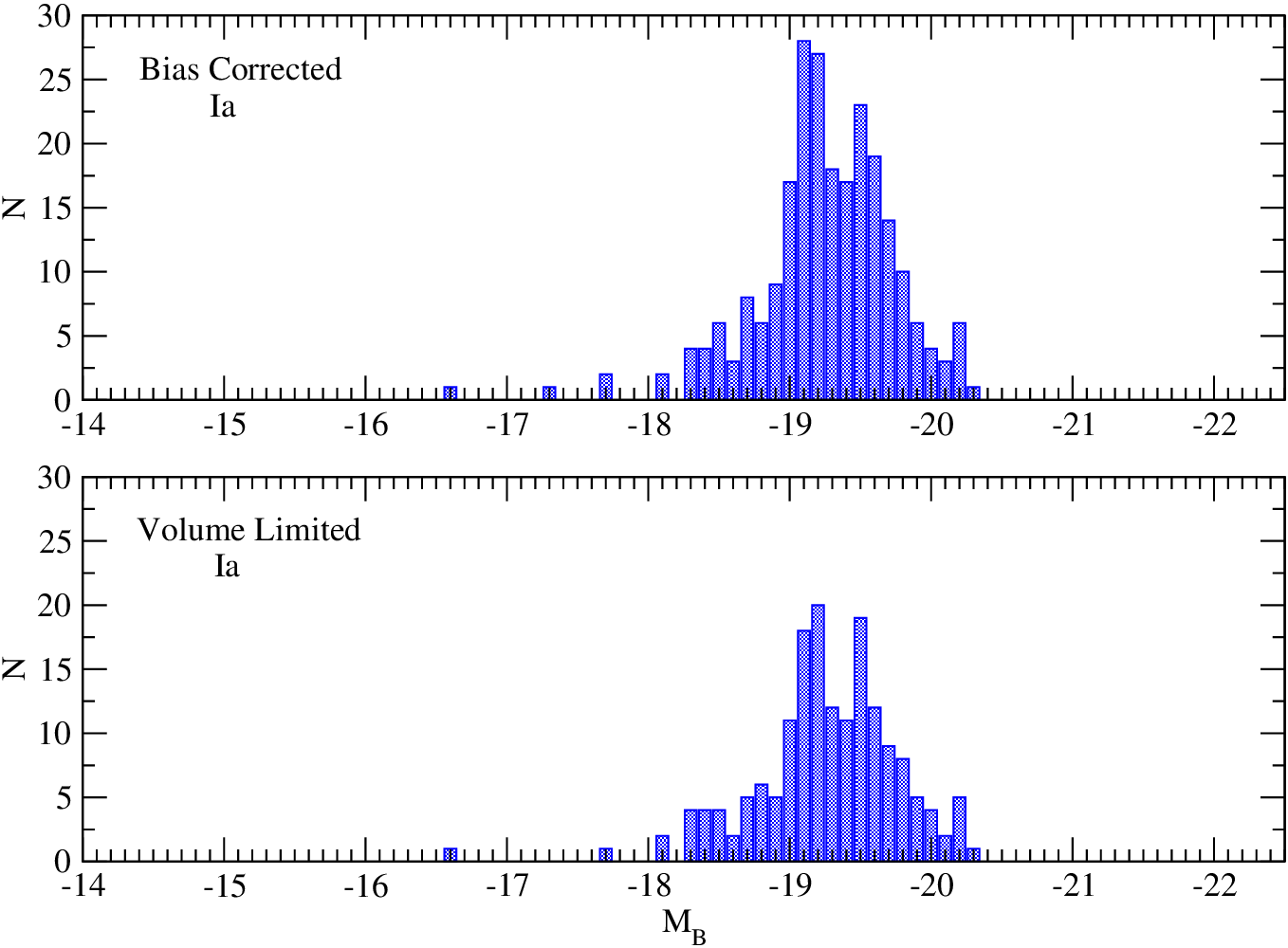}
\caption{\label{fig5}
Absolute--magnitude distributions for normal SNe~Ia. The top panel displays the bias--corrected sample,  
while the bottom panel displays the volume--limited sample.  
}
\end{figure} 

\clearpage 

\begin{figure}
\plotone{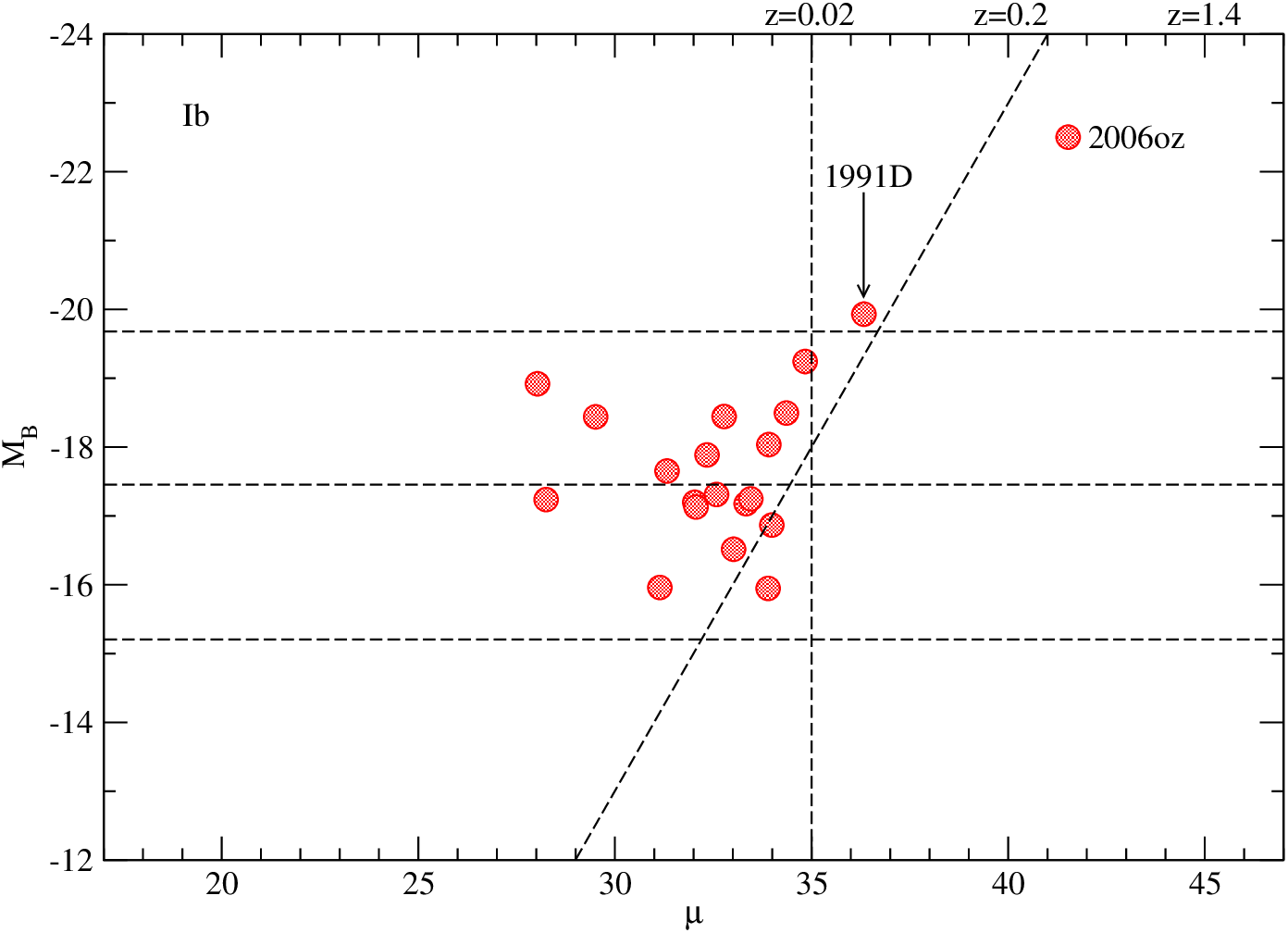} 
\caption{\label{fig6}
Miller diagram of 20 SNe~Ib. 
}
\end{figure} 

\begin{figure}
\plotone{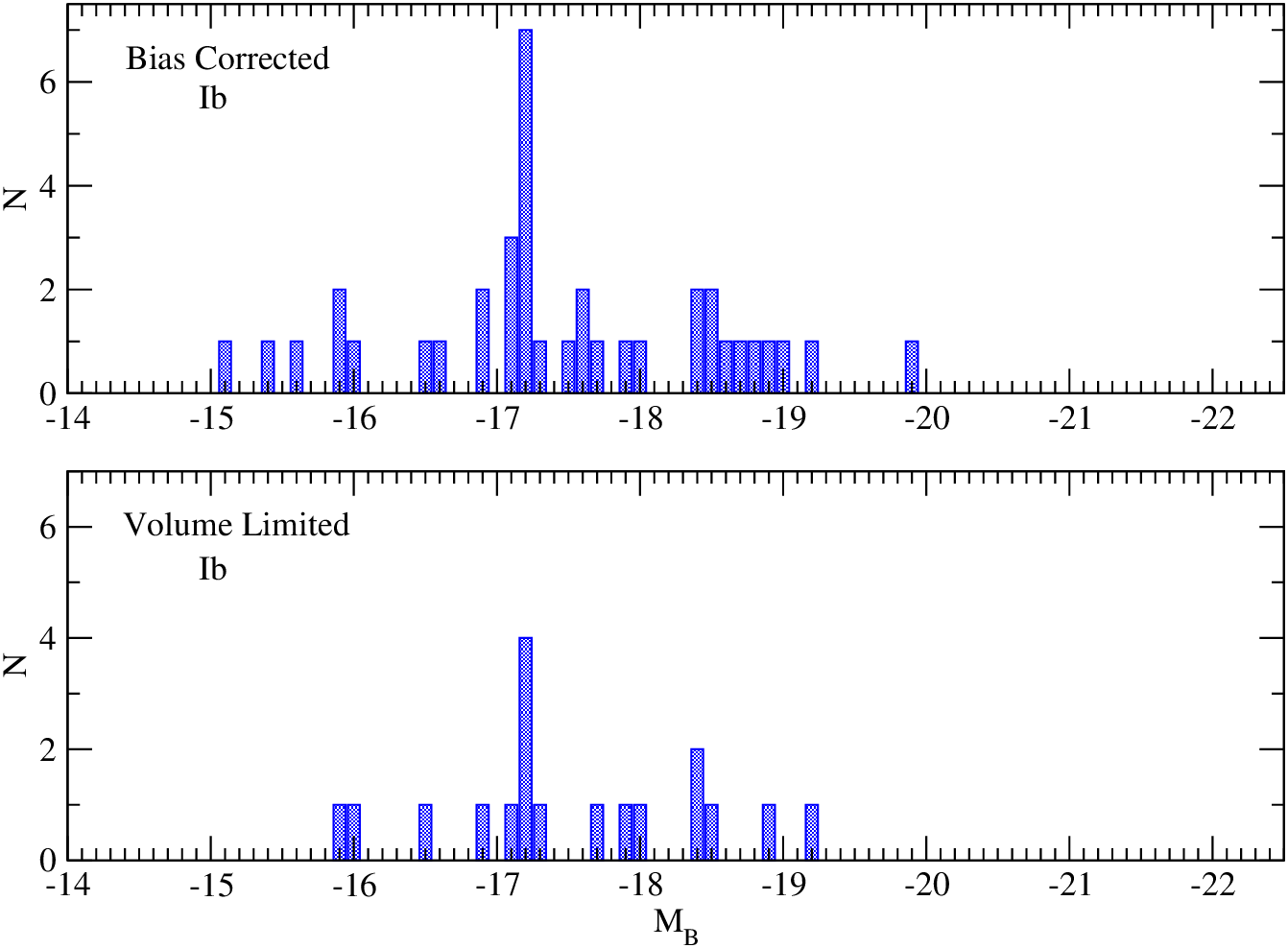}
\caption{\label{fig7}
Absolute--magnitude distributions for SNe~Ib. 
}
\end{figure} 

\clearpage 

\begin{figure}
\plotone{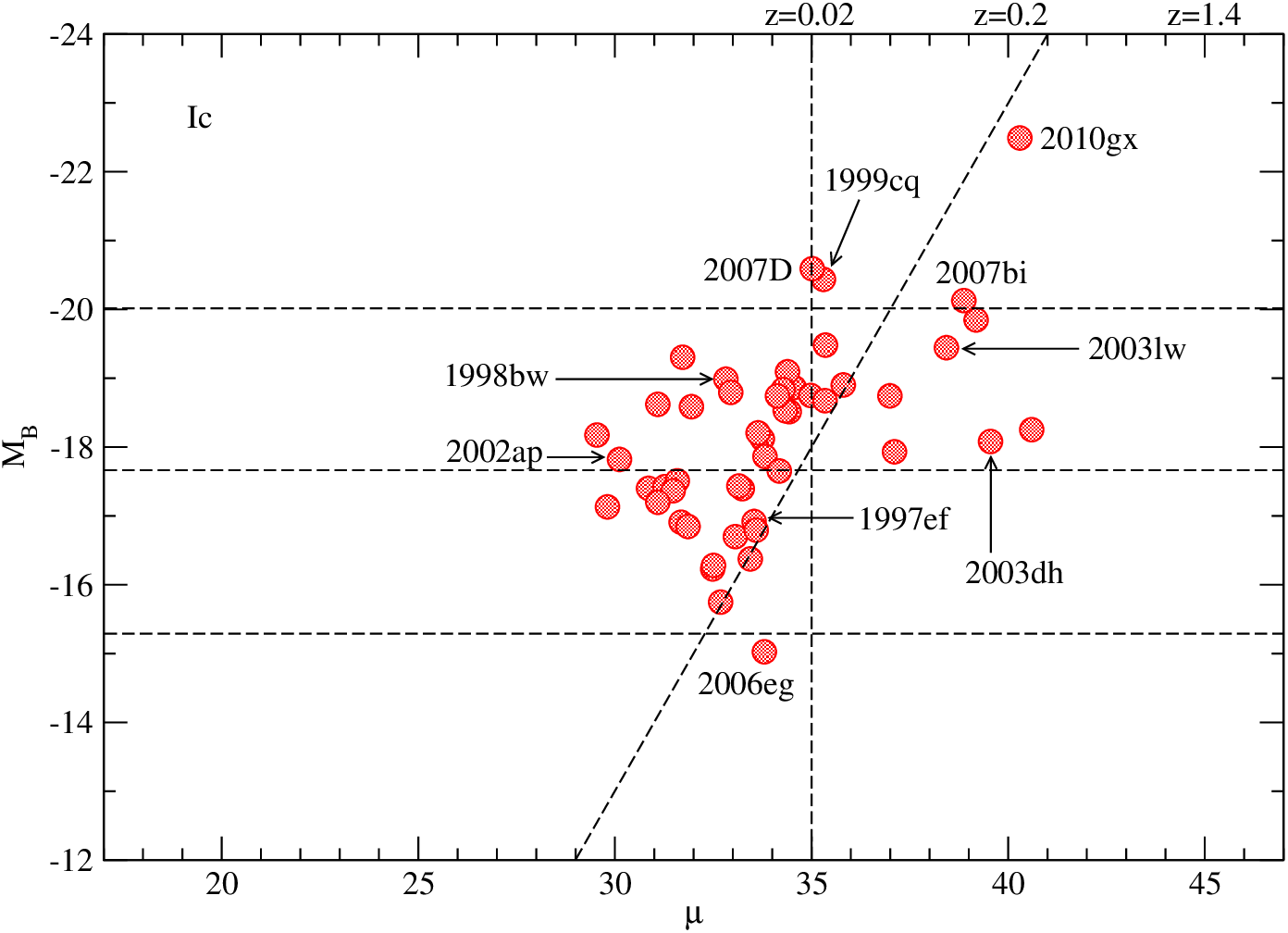} 
\caption{\label{fig8}
Miller diagram of 49 SNe~Ic. Along with the extreme SNe, several noteable SNe~IcBL are labeled as well.  
}
\end{figure} 

\begin{figure}
\plotone{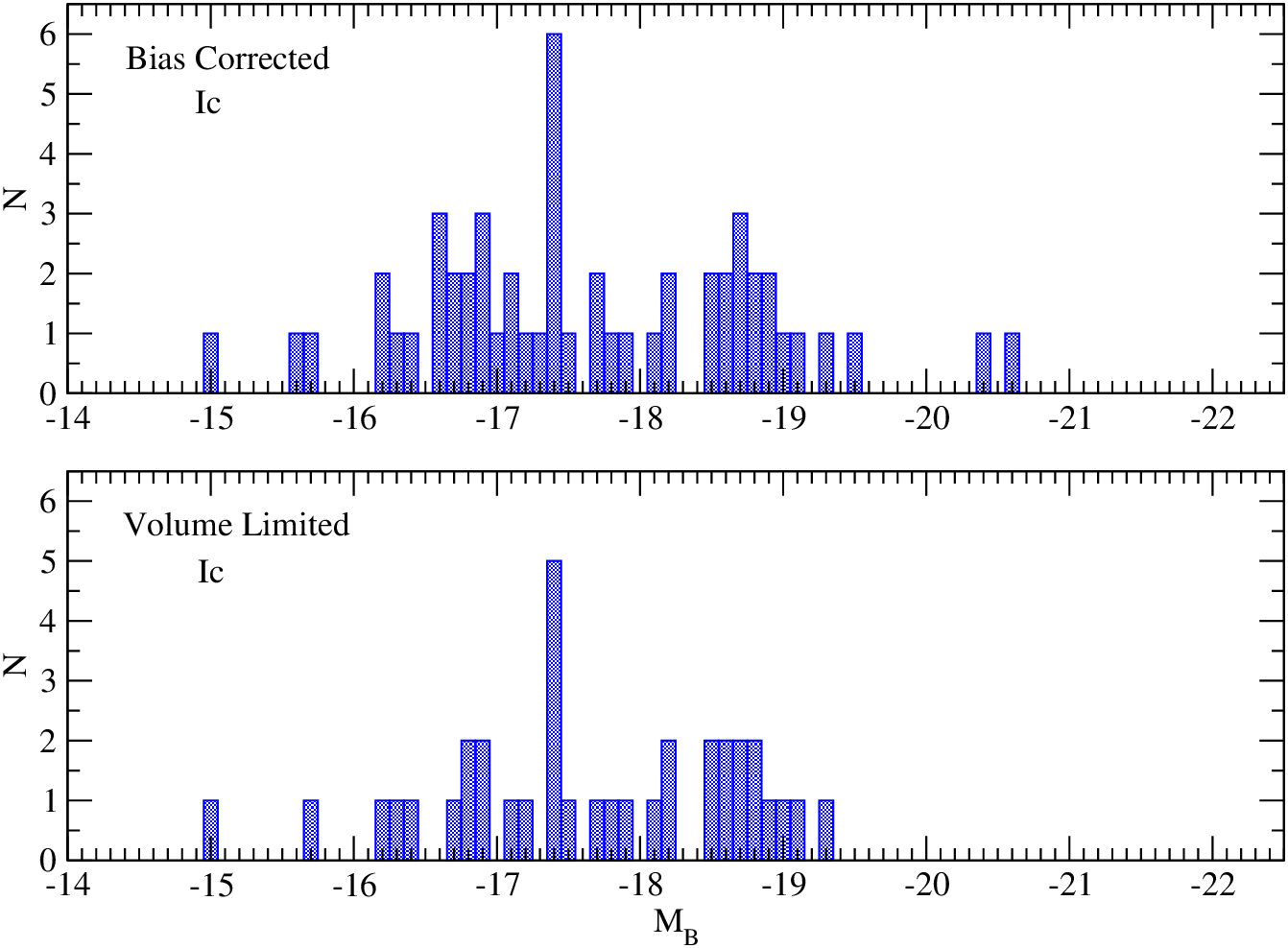}
\caption{\label{fig9}
Absolute--magnitude distributions for SNe~Ic. 
}
\end{figure} 

\clearpage 

\begin{figure}
\plotone{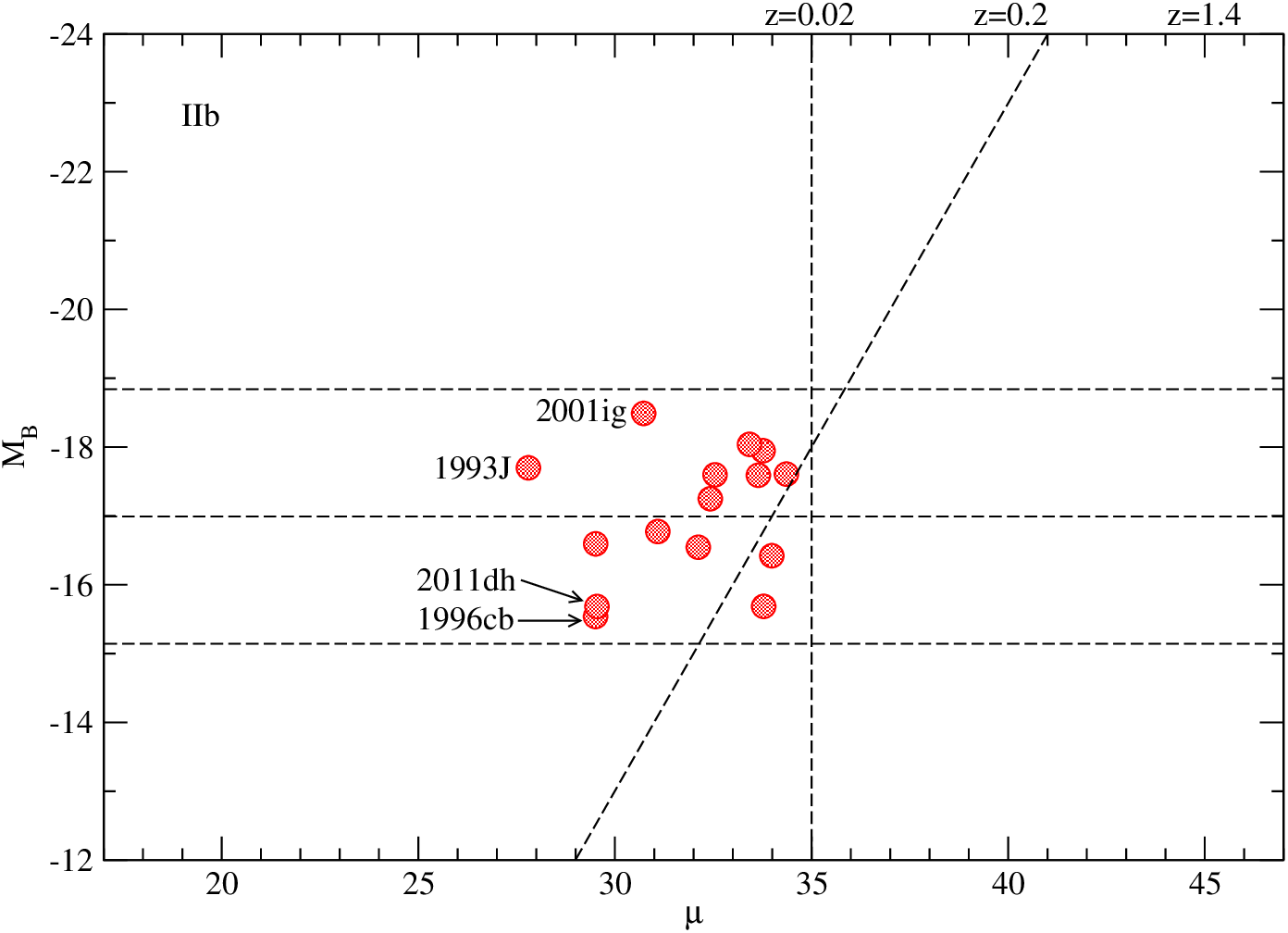} 
\caption{\label{fig10}
Miller diagram of 15 SNe~IIb. 
}
\end{figure} 

\begin{figure}
\plotone{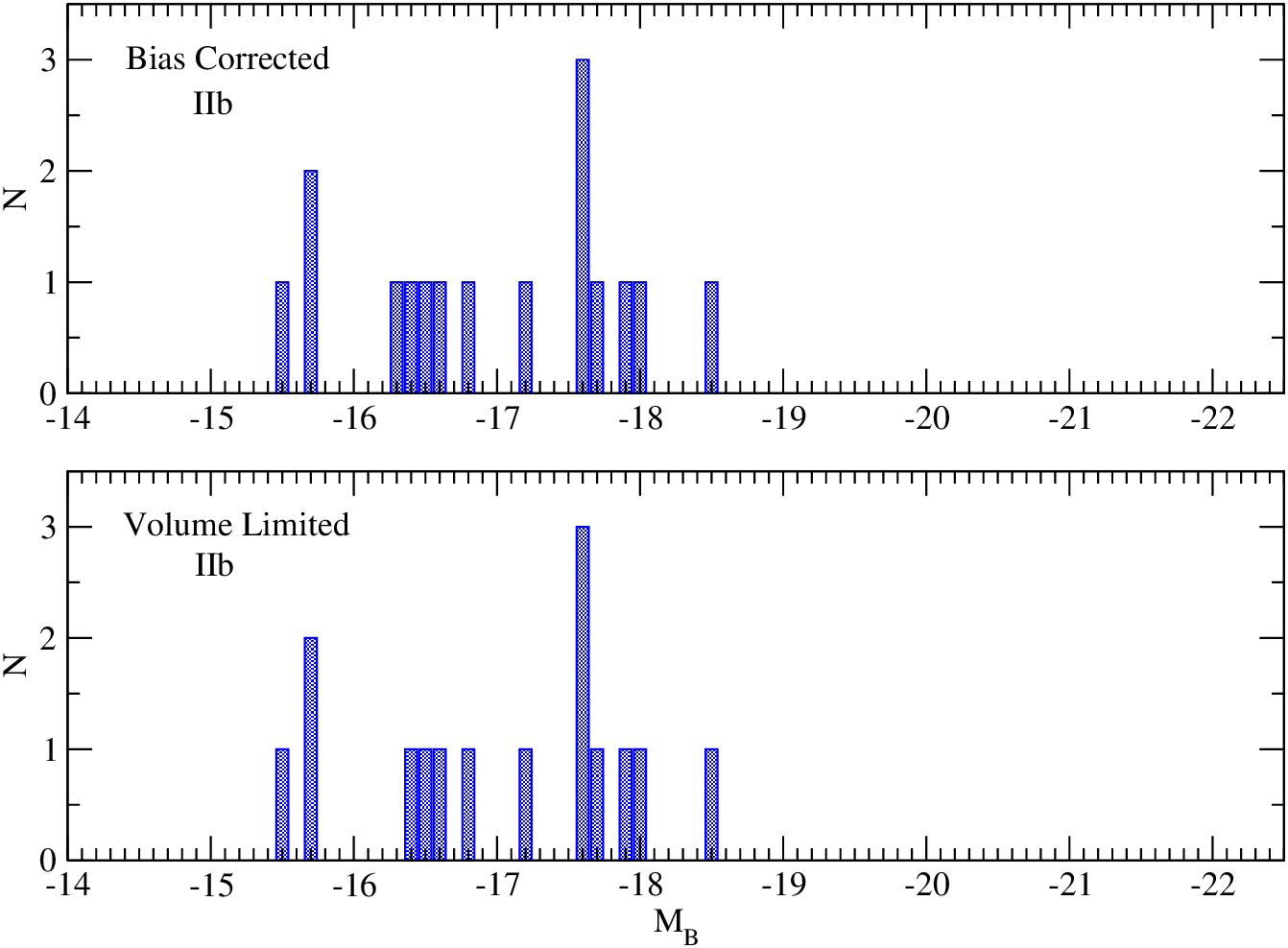}
\caption{\label{fig11}
Absolute--magnitude distributions for SNe~IIb.
}
\end{figure}

\clearpage 

\begin{figure}
\plotone{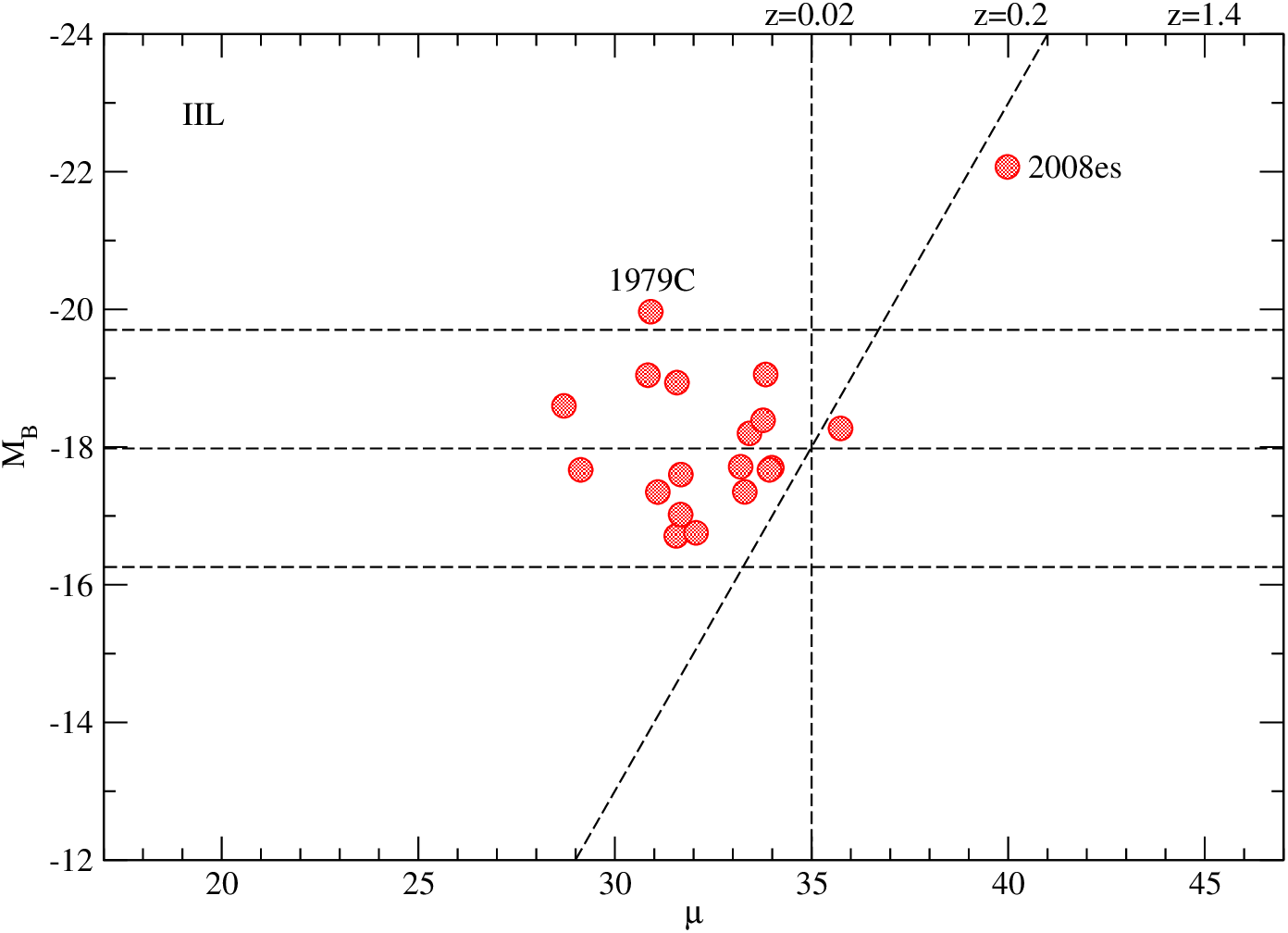} 
\caption{\label{fig12}
Miller diagram of 19 SNe~IIL. 
}
\end{figure}

\begin{figure}
\plotone{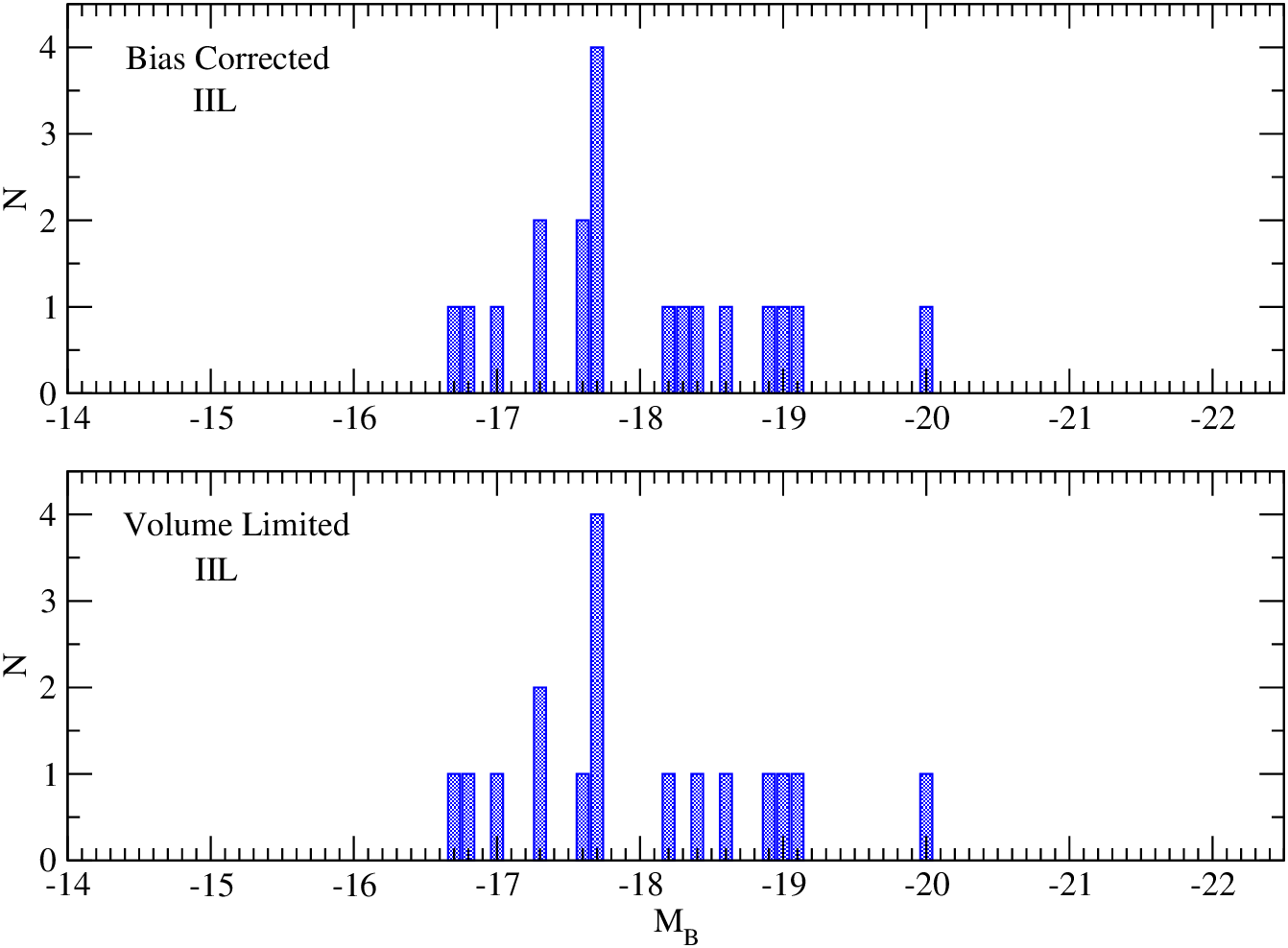}
\caption{\label{fig13}
Absolute--magnitude distributions for SNe~IIL. 
}
\end{figure} 

\clearpage 

\begin{figure}
\plotone{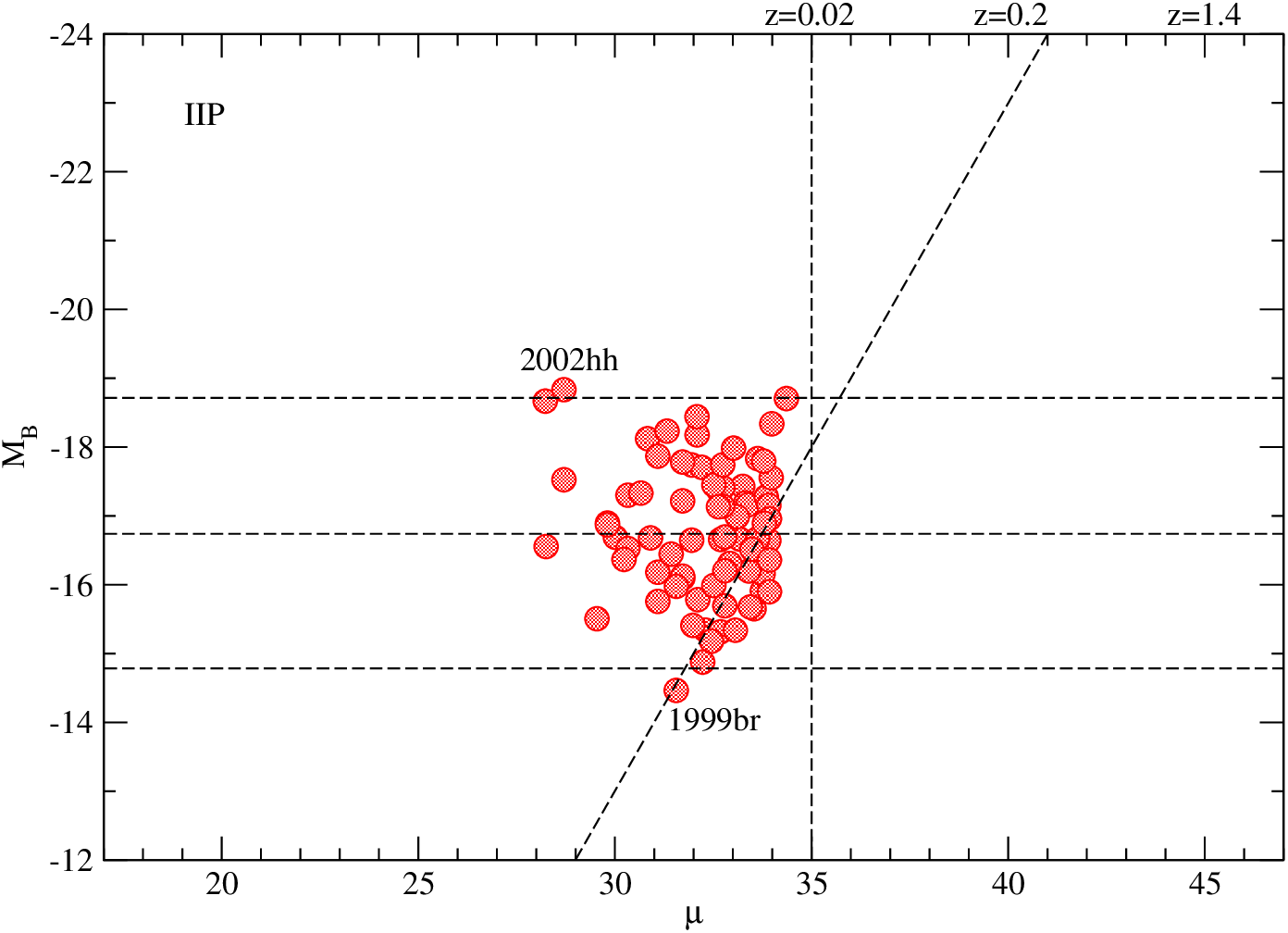} 
\caption{\label{fig14}
Miller diagram of 74 SNe~IIP. 
}
\end{figure}

\begin{figure}
\plotone{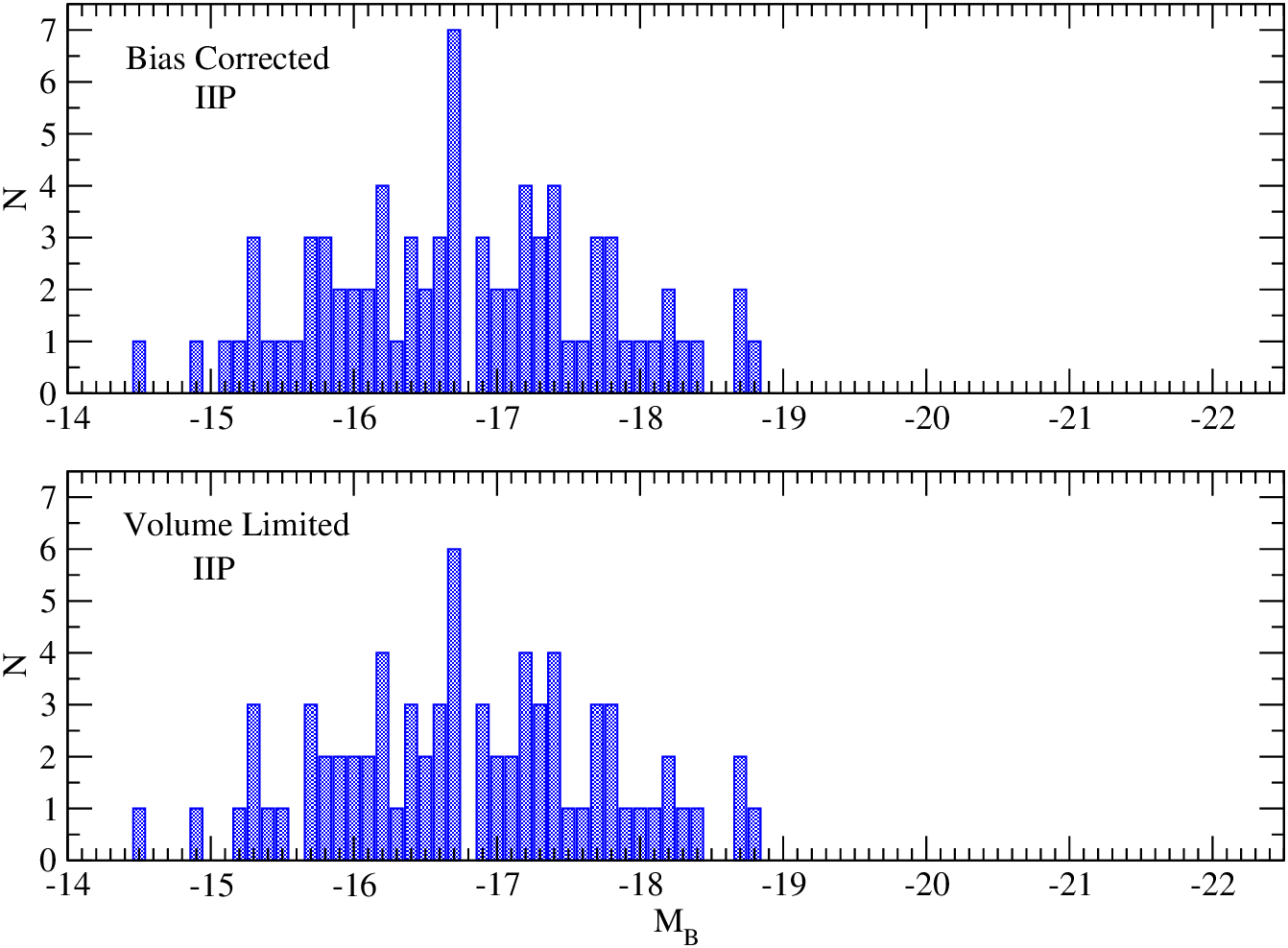}
\caption{\label{fig15}
Absolute--magnitude distributions for SNe~IIP. 
}
\end{figure} 

\clearpage 

\begin{figure}
\plotone{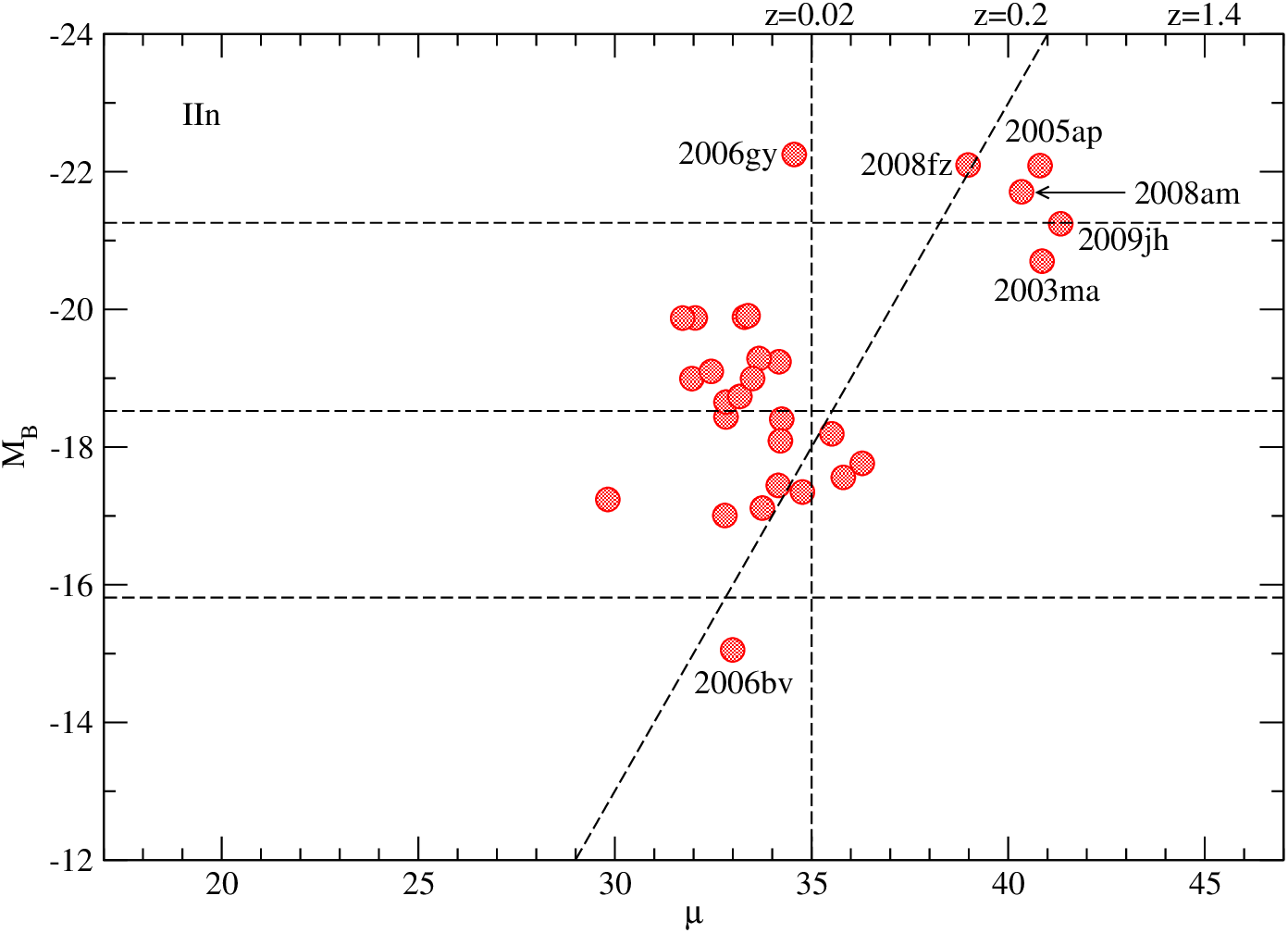}
\caption{\label{fig16}
Miller diagram of 29 SNe~IIn. 
}
\end{figure}

\begin{figure}
\plotone{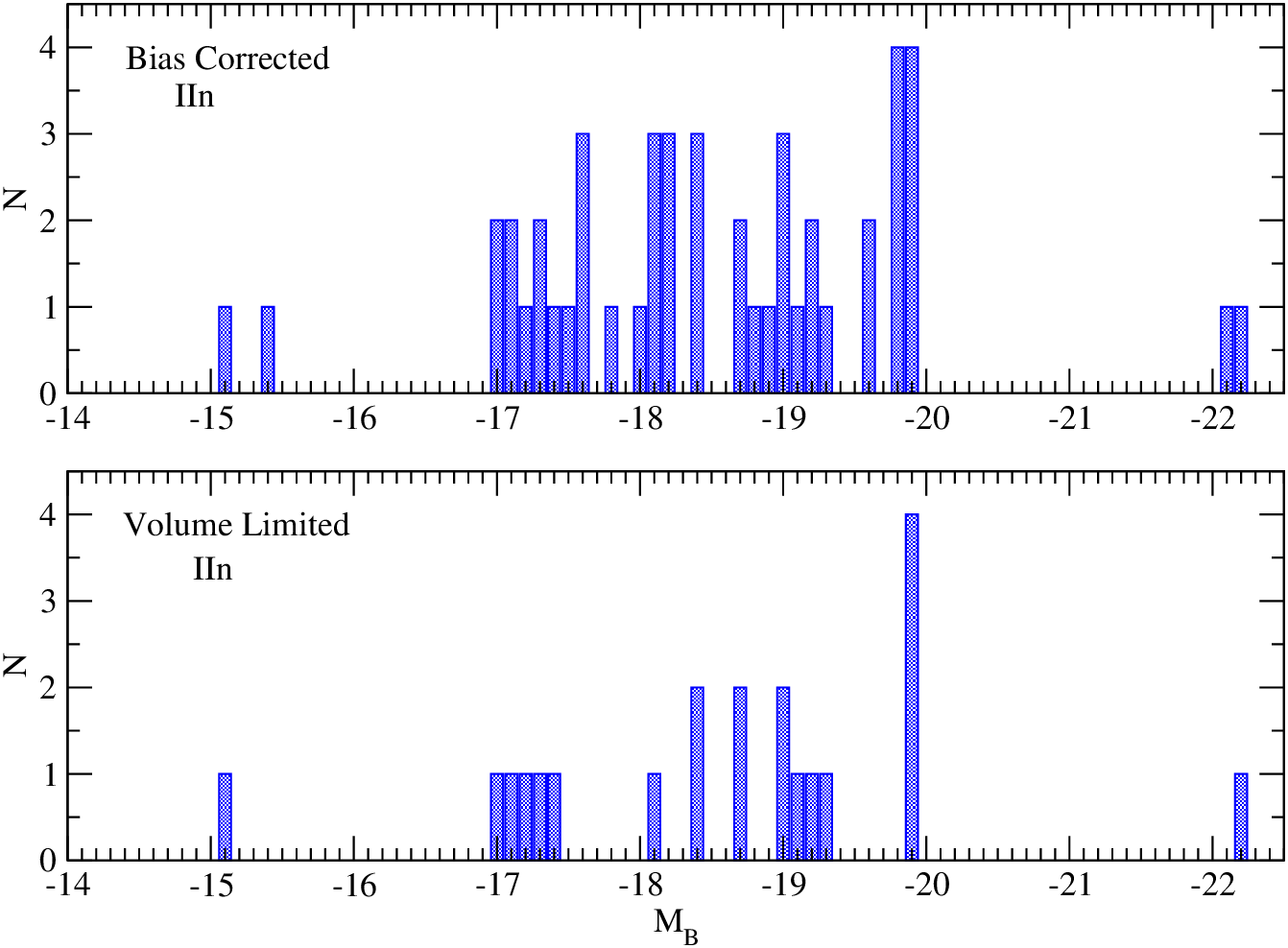}
\caption{\label{fig17}
Absolute--magnitude distributions for SNe~IIn.
}
\end{figure}

\end{document}